\documentclass{ptephy}
\usepackage{bm}
\usepackage{amsmath,amssymb,graphicx}

\newcommand{\Slash}[1]{{\ooalign{\hfil/\hfil\crcr$#1$}}}

\graphicspath{{./figM/}{./figVw/}{./figVL/}}

\begin{document}

\title{Weak solution of the non-perturbative renormalization group equation
to describe the dynamical chiral symmetry breaking}

\author{\name{Ken-Ichi Aoki}{\ast}, \name{Shin-Ichiro Kumamoto}{\ast} and \name{Daisuke Sato}{\ast}
}


\address{\affil{}{Institute for Theoretical Physics, Kanazawa University, Kanazawa 920-1192, Japan}
\email{aoki@hep.s.kanazawa-u.ac.jp; kumamoto@hep.s.kanazawa-u.ac.jp; satodai@hep.s.kanazawa-u.ac.jp}}

\begin{abstract}%
We analyze the dynamical chiral symmetry breaking (D$\chi$SB) in the Nambu-Jona-Lasinio (NJL) model 
by using the non-perturbative renormalization group (NPRG) equation.
The equation takes a form of two-dimensional 
partial differential equation for the multi-fermion effective interactions $V(x,t)$ where 
$x$ is $\bar\psi\psi$ operator and $t$ is the logarithm of the renormalization scale.
The D$\chi$SB occurs due to the quantum corrections, which means
it emerges  at some finite $t_{\rm c}$ in the mid of integrating the equation
with respect to $t$. At $t_{\rm c}$ some singularities suddenly appear
in $V$ which is compulsory in the spontaneous symmetry breakdown.
Therefore there is no solution of the  equation beyond $t_{\rm c}$.
We newly introduce the notion of weak solution to get the global solution including the
infrared limit $t\rightarrow \infty$ and investigate its properties. 
The obtained weak solution is global and unique,  and it perfectly
describes the physically correct vacuum even in case of the first order phase
transition appearing in finite density medium. 
The key logic of deduction is that the weak solution we defined 
automatically convexifies the effective potential when treating the singularities.
\end{abstract}

\subjectindex{B00, B02, B32, B69}

\maketitle

\section{Introduction}

The dynamical chiral symmetry breaking (D$\chi$SB) has been the central issue of the elementary
particle physics since it was initially founded by Nambu and Jona-Lasinio
\cite{Nambu:1961tp, Nambu:1961fr}.
In short, all the variety of elementary particles comes out of the special pattern of 
the spontaneous chiral symmetry breaking of some unified gauge theory.
For the quarks, the strong interactions described by QCD breaks the chiral symmetry
at the infrared scale, where the QCD gauge coupling constant becomes enough strong.
Recently QCD in high temperature and density has drawn strong attention since
there are fruitful new phases expected, including the chiral symmetry restoration and the color
superconductivity.

The D$\chi$SB is highly non-perturbative phenomena and
cannot be treated by the perturbation theory. Thus there have been various
non-perturbative approaches to analyze it
such as the lattice simulation and the Schwinger-Dyson (SD) approaches.

Here we use the non-perturbative renormalization group (NPRG)  method 
originated from the Kadanoff and Wilson's \cite{Wilson:1973jj,Kadanoff:1966wm} idea.
The NPRG approaches share good features; 
it has no sign problem at finite chemical potential, which has suffered the lattice simulation 
seriously; it can improve the approximation systematically so as to reduce 
the gauge dependence of physical quantities \cite{Aoki:2012mj,Aoki:2000dh}, 
which is the essential problem of the SD approaches.

Now we state the central subject of this paper. 
In the NPRG method, we integrate the path integral of the theory  
from the micro scale to the macro scale, slice by slice, and obtain the differential equation
of the effective action with respect to the logarithmic renormalization scale $t$, which is  
defined by
\begin{equation}
t =-\log(\Lambda/\Lambda_0),
\label{defoft}
\end{equation} 
where $\Lambda_0$ is the bare cut off scale and $\Lambda$ is the renormalization scale.

In the simplest approximation, the effective action is expressed in terms of the field
polynomials without derivatives in addition to the normalized kinetic terms. Then 
the coupling constants $C_i$, the coefficients of these operator polynomials,
are functions of $t$. The change of these coupling constants are given by 
functions of the coupling constants themselves,
\begin{equation}
\frac{d}{dt} C_i(t) = \beta_i\left(C(t)\right),
\end{equation}   
where the right-hand side is called the $\beta$-function.
This differential equation is the renormalization group equation (RGE).

The $\beta$-function is evaluated by the one-loop diagrams only and the propagators 
constituting the loop are limited to have the sliced momentum. Therefore its
evaluation is so simple that only the angular integration should be done to leave the
total solid angle factor, which is called the shell mode integration.

Here we consider the four-fermi coupling constant $G$. The four-fermi interactions play
an essential role in the total framework of the D$\chi$SB analysis.
The $\beta$-function for the {\sl dimensionless} four-fermi coupling constant 
rescaled by the renormalization scale $\Lambda$, $\tilde G(\equiv \Lambda^2 G)$, 
is given by the following
Feynman diagrams,
\begin{align}
\frac{d}{dt}\tilde G&=-2 \tilde G +\ \ 
\begin{minipage}[c]{0.5\hsize}
\includegraphics[scale=1.0]{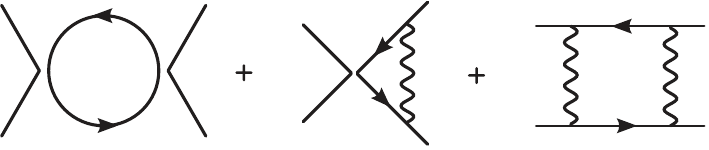}
\end{minipage} \!\!,
\end{align}
where the first term in the right-hand side represents the canonical scaling due to the
operator dimension, the solid lines are quarks, the wavy lines are gluons and 
the above loops represent the shell mode integration. 
It is evaluated as follows \cite{Aoki:1998gt,Aoki:1999dw,Aoki:1999dv},
\begin{align}
\frac{d}{dt}\tilde G &= -2\tilde G+\frac{1}{2\pi^2}\left(\tilde G+(3+\xi)C_2 \pi \alpha_{\rm s}\right)^2,
\end{align} 
where $\alpha_{\rm s}$ is the gauge coupling constant squared as usual, $C_2$ is 
the second Casimir invariant of the quark representation and  $\xi$ is the
gauge fixing parameter in the standard $R_\xi$ gauge.  
Due to the chiral symmetry, the RGE for four-fermi operator is closed by itself, no contribution 
from higher dimensional operators allowed.

\begin{figure}
\centering
\includegraphics[width=0.4\hsize]{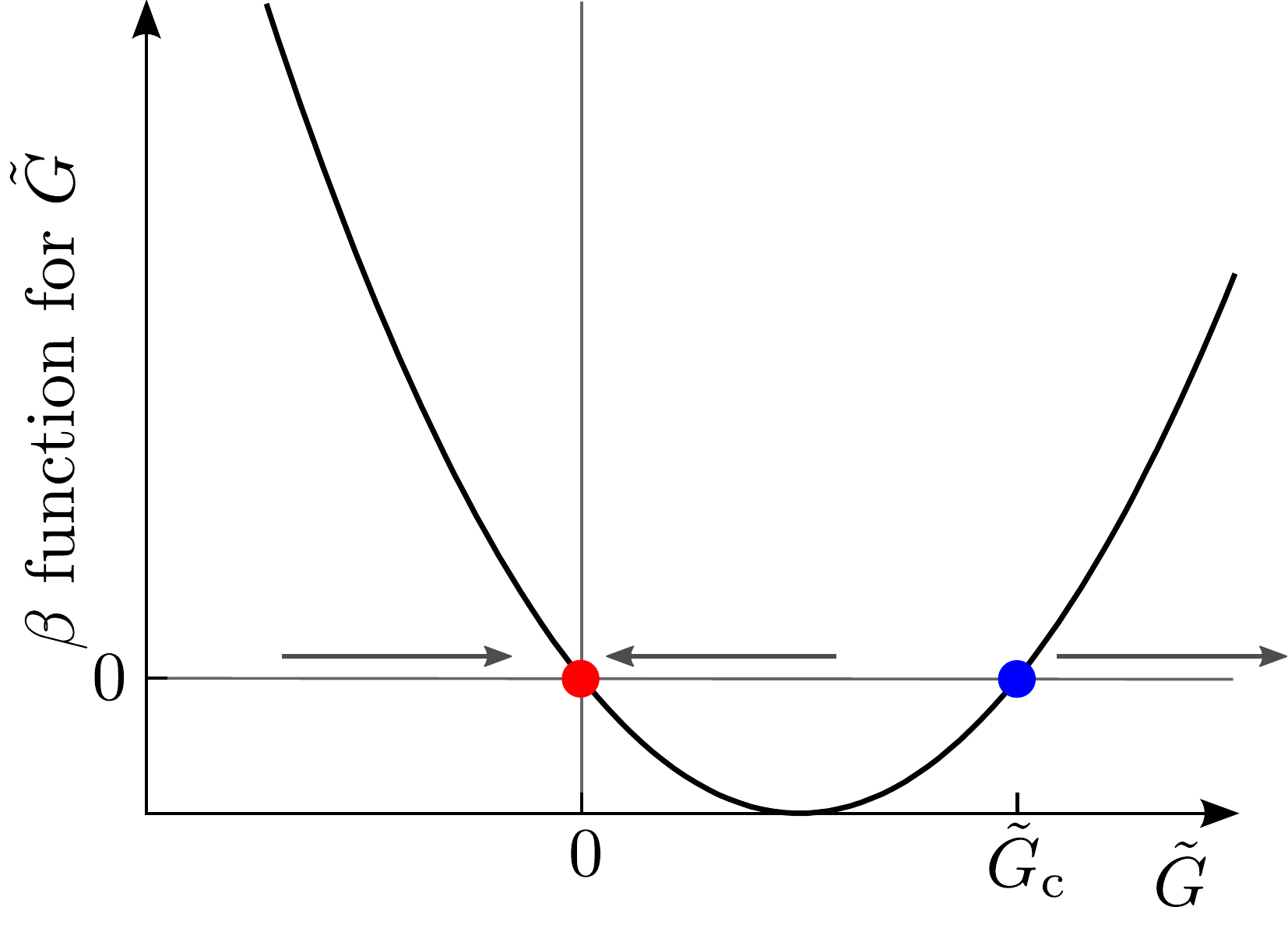}
\caption{The $\beta$ function for the four-fermi coupling constant $\tilde G$ in the NJL model. }
\label{fig:NJLbeta}
\end{figure}

Consider first the Nambu-Jona-Lasinio (NJL) model as an effective infrared model of QCD.
Then the $\beta$-function is simplified to give
\begin{align}
\frac{d}{dt} \tilde G &= -2\tilde G+\frac{1}{2\pi^2}\tilde G^2,
\label{eq:NJLbeta}
\end{align}
and it is depicted in Fig.\,\ref{fig:NJLbeta}, where the arrows show the direction of 
renormalization group flows towards the infrared.
Note that there are two fixed points, zeros of the $\beta$-function, 
at 0 and  $\tilde G_{\rm c}=4\pi^2$, and
they are infrared (stable) fixed point and ultraviolet (unstable) fixed point respectively.

It is readily seen that there are two phases, the strong phase and the weak phase, and
the critical coupling constant dividing the phase is $\tilde G_{\rm c}$. 
If the initial coupling constant is larger than $\tilde G_{\rm c}$, flows go to the positive
infinite, otherwise they approaches to the origin.
We are sure that this is the quickest argument ever to derive the critical coupling constant
in the NJL model and the result is equal to the mean field approximation.

The weak phase has no problem here. The flows approaching to the origin do not mean that the
four-fermi interactions vanish at the infrared. Since the variable here is the dimensionless
coupling constant, the four-fermi interactions look vanishing due to the multiplication 
factor of $\exp(-2t)$. As for the dimensionful variable, it just stops running to give some
constant size infrared interactions determined by the initial coupling constant.

\begin{figure}
\centering
\includegraphics[width=0.35\hsize]{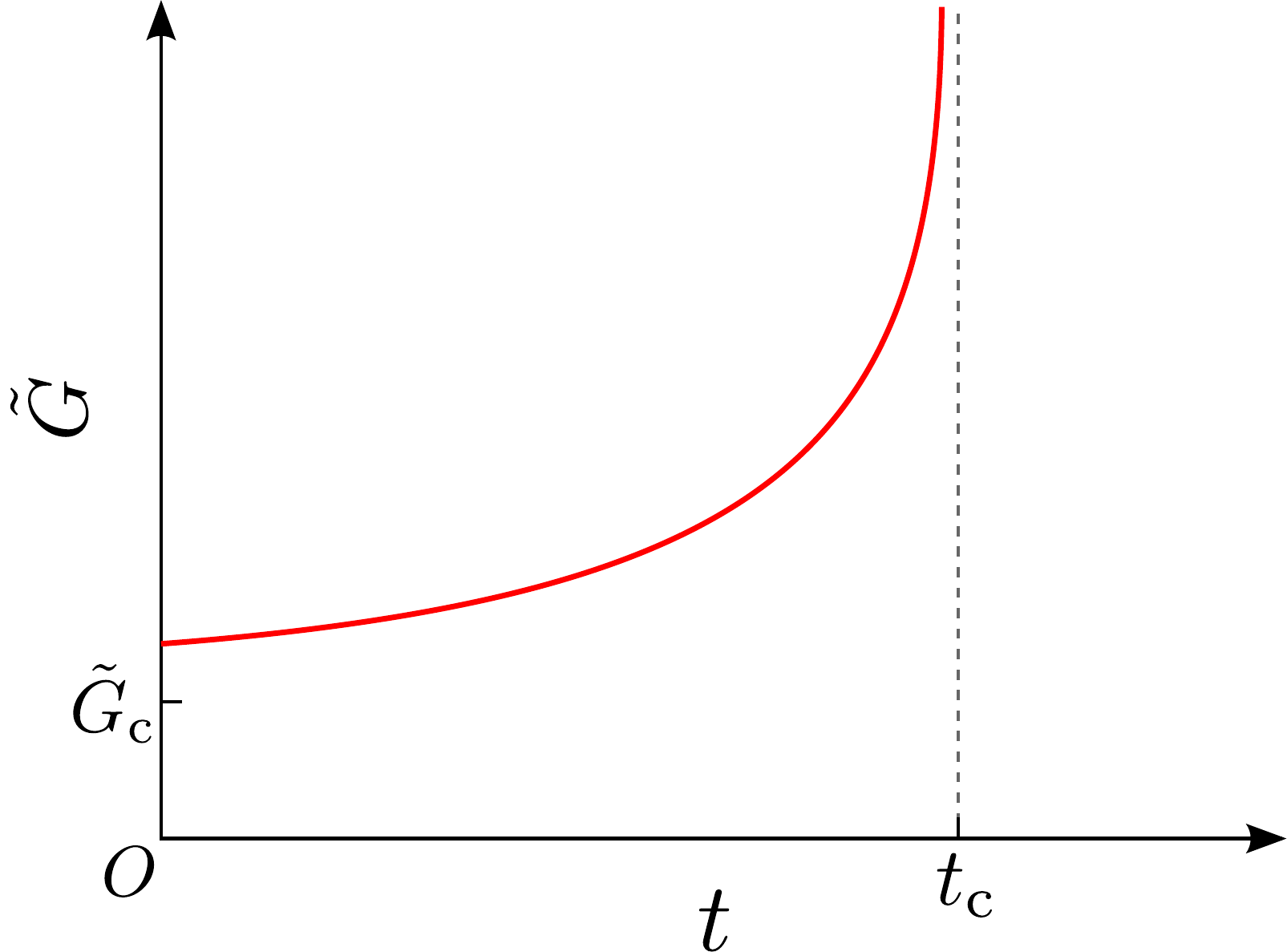}
\caption{Blowup behavior of the running four-fermi coupling constant $\tilde G$.}
\label{NJLblowup}
\end{figure}

On the other hand, the strong phase has a serious problem.
Equation\,(\ref{eq:NJLbeta}) is easily integrated to give the following solution,
\begin{align}
\tilde G(t) = \frac{\tilde G_{\rm c}\tilde G_0}{\tilde G_0-\left(\tilde G_0-\tilde G_{\rm c}\right){\rm e}^{2t}} ,
\end{align}
where $\tilde G_0$ is the initial coupling constant (Fig.\,\ref{NJLblowup}). This solution is a blowup solution
and it diverges at a finite $t_{\rm c}$ defined by,
\begin{align}
t_{\rm c}= \frac{1}{2}\log\frac{\tilde G_0}{\tilde G_0-\tilde G_{\rm c}}.
\end{align}
Because of this blowup behavior, we cannot obtain the solution after $t_{\rm c}$, that is, there is 
no global solution in Eq.\,(\ref{eq:NJLbeta}) for $\tilde{G}_{\rm 0}>\tilde{G}_{\rm c}$.

This is not a fake nor an insufficientness of our treatment.
This must happen necessarily \cite{Aoki:2000wm,Braun:2011pp}. 
The reason is the following. The NJL model is expected to have the D$\chi$SB
with strong coupling 
constant $\tilde G_0 > \tilde G_{\rm c}$. 
This is the quantum effects, that is, it is caused by the quantum loop corrections.
The NPRG method calculates the effective interactions by adding the quantum
loop corrections from micro to macro, step by step. Then the initially symmetric
theory shows up quite a different effective interactions exhibiting the spontaneous
mass generation, at some intermediate stage of integration of the NPRG equation.

\begin{figure}
\centering
\includegraphics[width=0.4\hsize]{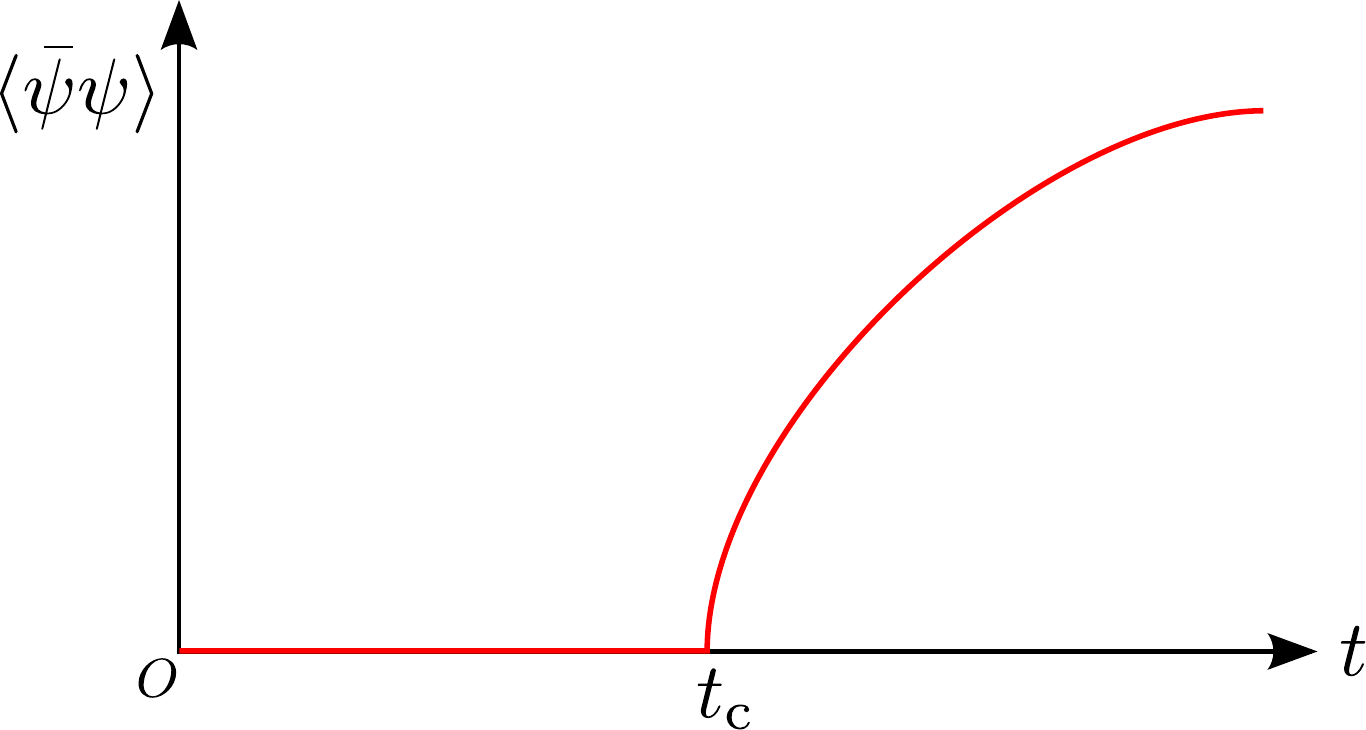}
\caption{Second order phase transition.}
\label{secondorderPT}
\end{figure} 

The schematic view of the chiral condensates or the dynamical mass as a function of the
renormalization scale $t$ is drawn in Fig.\,\ref{secondorderPT}, 
where at $t_{\rm c}$ the D$\chi$SB occurs.
This figure resembles the {\sl temperature} dependence of the 
spontaneous {\sl magnetization} if the low and high $t$ are reversed. 
Then consider what happens if $t$ approaches $t_{\rm c}$ from the origin.
The correlation length of  {\sl spins} ($\bar\psi\psi$ operators) must diverge and also 
the {\sl magnetization} susceptibility diverges. The susceptibility is proportional to the total 
{\sl spin} fluctuation  $\langle (\bar\psi\psi)^2\rangle$. This is nothing but the 
four-fermi interactions. Therefore the diverging behavior of the four-fermi interactions itself
is a normal and naive phenomenon that just tells us we are approaching the second order
phase transition point. 

\begin{figure}[h]
\centering
\includegraphics[width=0.5\hsize]{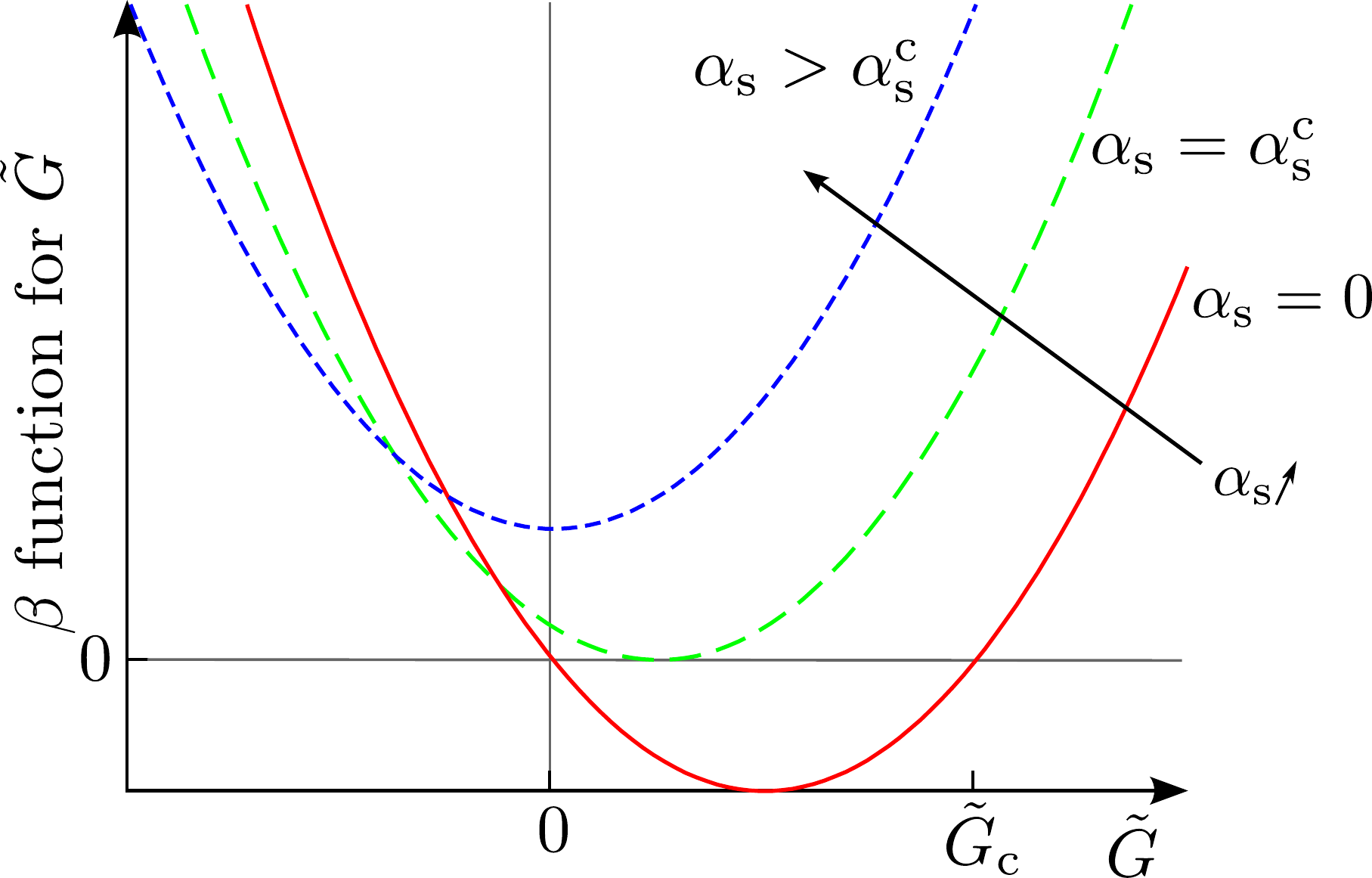}
\caption{The $\beta$ function for the four-fermi coupling constant $\tilde G$ in gauge theories.}
\label{QCDbeta}
\end{figure}
  
\begin{figure}
\centering
\includegraphics[width=0.45\hsize]{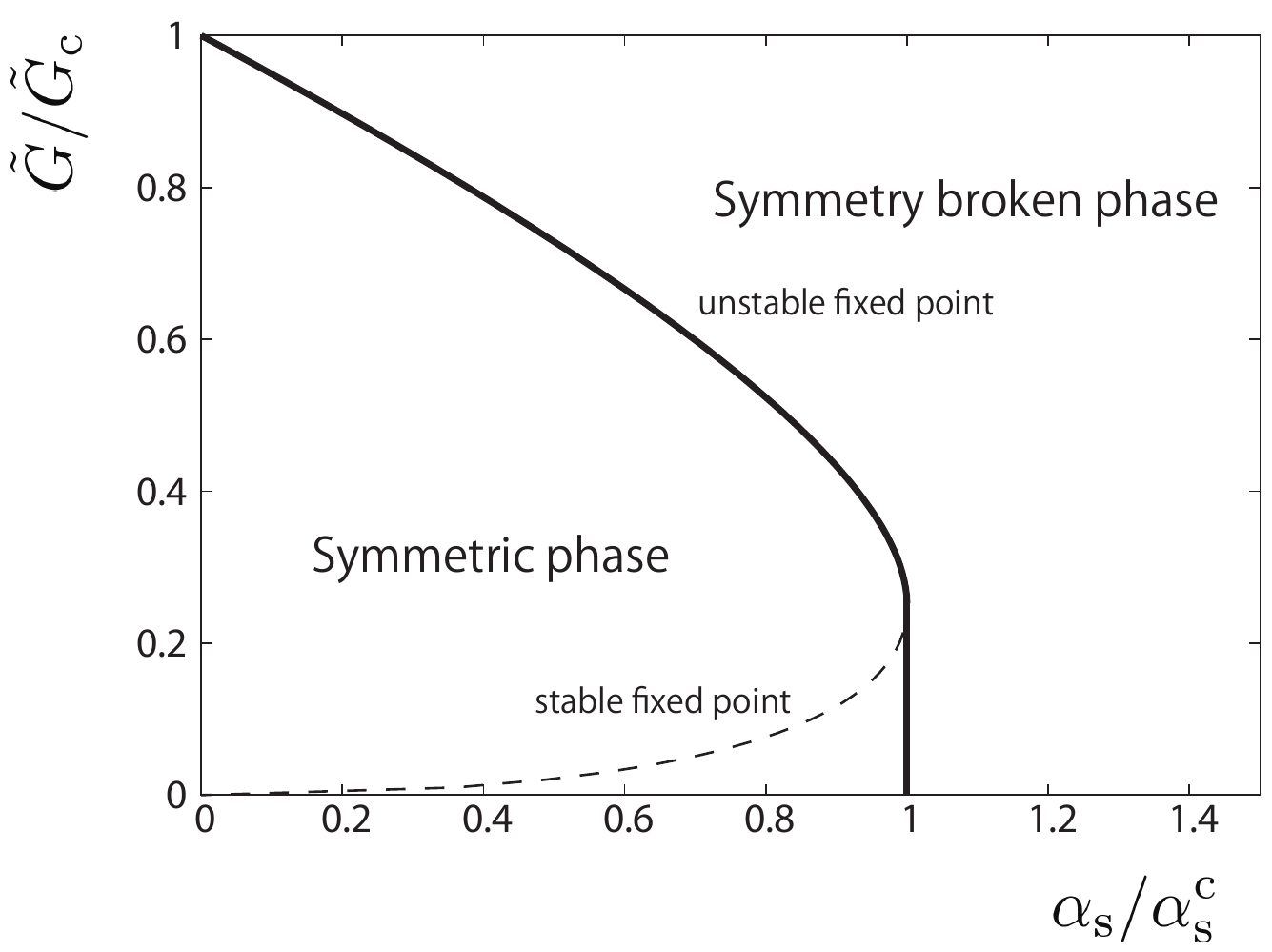}
\caption{Phase structure of strong QED in $\alpha_{\rm s} - \tilde G$ plane.}
\label{sqed}
\end{figure}

The same story holds for QCD case. The $\beta$-function of the four-fermi coupling
constant now depends on the gauge coupling constant as shown in Fig.\,\ref{QCDbeta}
 ($\xi=0, C_2=1$ case).
Switching on the gauge interactions, the gauge coupling constant becomes large towards the
infrared, the parabola function moves up, and finally 
after $\alpha_{\rm s}>\alpha_{\rm s}^{\rm c}=\pi/3$, it has no
zero at all. 
Accordingly, the stable and
unstable fixed points approach to each other, and they are merged to 
pair-annihilate. After this annihilation, total space belongs to the strong phase and the four-fermi
coupling constant moves to positive infinite. 
This is our NPRG view of how QCD breaks the chiral symmetry in the
 infrared, and also it explains the peculiar phase diagram of 
the so-called strong QED (or the gauged NJL model) as is depicted in 
Fig.\ref{sqed} from the RGE point of view\cite{Aoki:1998gt,Aoki:1999dw,Aoki:1999dv}.

Now the problem is how to calculate the infrared quantities, such as the chiral condensates and 
the dynamical mass. They reside all beyond the blowup point.
We stress here that this blowup is irrelevant to the approximation scheme.
Any theory, approximated or not, as far as it exhibits 
the D$\chi$SB,  must encounter
the spontaneous breaking in the mid of the renormalization, since it is produced  
by the sufficient quantum corrections. 
The D$\chi$SB must be expressed by the singular 
effective interactions of fermions.
Therefore, any model providing the D$\chi$SB 
must meet singularities in the mid of
renormalization. This is inevitable.

Many NPRG analyses of the D$\chi$SB  have been performed 
by introducing the bosonization of the multi-fermi interactions
\cite{Aoki:1996gx,Aoki:1998gt,Aoki:1999dw,Gies:2001nw}, which is called the
auxiliary field method or the Hubbard-Stratonovich transformation.
This type of approaches may avoid the blowup singularity
and  has worked well indeed. However it must contain additional 
difficulties or ambiguities 
in how to treat and approximate the system of bosonic degrees of freedom.

We take the other way around. 
We do not introduce the bosonic degrees of freedom and instead
we extend the notion of the solution of NPRG equation. 
First we go back to  the original NPRG equation 
without expansion in polynomials of field operators.
Then it is a partial differential equation (PDE) of at least two variable, the operator and 
renormalization scale $t$.
The blowup behavior is reexamined in the next section 
and it is made clear that even the solution of PDE must encounter the corresponding
singularity at $t=t_{\rm c}$. Thus the origin of singularity is irrelevant to the Taylor
expansion itself of PDE.
Even if the PDE type of RGE is adopted, we cannot have any global
solution.

Then we introduce 
the notion of weak solution of PDE \cite{Evans:2010}.
It allows solution with singularities and we may define the global
solution including the infrared limit.

This paper is organized as follows. 
In Sect.\,2, taking the NJL model as an example, 
we briefly review the Wegner-Houghton equation, an approximation of
NPRG equation.
In Sect.\,3, the notion of weak solution is introduced and 
a practical method to construct the
weak solution is explained in Sect.\,4.
In Sect.\,5, we solve the NJL model to explicitly construct the weak solution.
In Sect.\,6, we discuss the effect of the bare mass of quarks, which is necessary 
to define the chiral order parameters, and we construct the Legendre effective potential. 
In Sect.\,7, the method of weak solution is applied to 
the first order phase transition case in the finite chemical potential NJL model. 
The convexity of the effective potential given by the weak solution is fully discussed.
Finally we summarize this paper in Sect.\,8.

To conclude, we will prove that the weak solution of NPRG equation is perfectly successful,
giving the global solution to calculate the infrared physical quantities.
Most impressive is the case of the first order phase transition.
There the weak solution correctly picks up the lowest free energy
vacuum among the multi locally stable vacua, through the procedure that
the effective potential is automatically convexified by the weak solution. 

   \section{Non-perturbative renormalization group}

In this section we introduce the non-perturbative renormalization group
by using the Wegner-Houghton (WH) equation.
For simplicity, we take the Nambu--Jona-Lasinio (NJL)  
model with a simplified discrete chiral symmetry, 
which is regarded as a low energy effective model of QCD explaining the D$\chi$SB. 

The Lagrangian density is given by
\begin{align}
\mathcal{L}&=\bar\psi \Slash{\partial} \psi 
-\frac{G_0}{2} (\bar\psi\psi)^2,
\label{eq:NJL}
\end{align}
where $\psi$ and $\bar\psi$ are the quark field and the antiquark field, respectively.
There is no continuous chiral symmetry. However, the Lagrangian 
is invariant under the following discrete chiral ($\gamma_5$) transformation,
\begin{align}
\psi \rightarrow \gamma_5\psi,\ \bar\psi\rightarrow -\bar\psi\gamma_5.
\label{eq:ChrlTrnsfmtn}
\end{align}
This discrete chiral symmetry forbids the mass term $m\bar\psi\psi$ and the chiral condensate $\langle\bar\psi\psi\rangle$ as well as the usual continuous chiral symmetry.

It is first clarified by Nambu and Jona-Lasinio \cite{Nambu:1961tp, Nambu:1961fr} that 
for the strong enough four-fermi coupling constant $G_0$, it exhibits D$\chi$SB.
Using the mean field approximation, 
the critical coupling constant $G_{\rm c}$ is found to be $4\pi^2/\Lambda_0^2$, 
where $\Lambda_0$ is the ultraviolet cutoff scale.
Note that the mean field approximation is equivalent to the self-consistency equation 
limited to the large-$N$ leading diagrams, where $N$ is the number of quark flavors.

In the NPRG approach, a central object is the Wilsonian effective action $S_{\rm eff}[\phi;\Lambda]$ 
defined by integrating out the microscopic degrees of freedom $\phi_{\rm H}$ with momenta higher than 
the renormalization scale $\Lambda$,
\begin{align}
\int\!\mathcal{D}\phi_{\rm H} e^{-S_0[\phi_{\rm L},\phi_{\rm H};\Lambda_0]}
= e^{-S_{\rm eff} [\phi_{\rm L};\Lambda]}, 
\label{eq:WHeq}
\end{align}   
where $S_0[\phi;\Lambda_0]$ is the bare action with the ultraviolet cutoff scale $\Lambda_0$ and 
$\phi$ generically denotes all relevant fields. 
The renormalization scale $\Lambda$ is parametrized by 
dimensionless variable $t$ as defined in Eq.\,(\ref{defoft}), 
\begin{align}
\Lambda(t) = \Lambda_0 e^{-t}.
\end{align}

The $t$-dependence of the effective action $S_{\rm eff}[\phi;t]$ is given by 
the NPRG equation, which is the following functional differential equation,
\begin{align}
\frac{d}{dt} S_{\rm eff} [\phi;t] =
 \beta_{\rm WH}\left[
   \frac{\delta S_{\rm eff}}{\delta \phi},\frac{\delta^2 S_{\rm eff}}{\delta \phi^2};t
 \right],
\end{align}
which is also called the Wegner-Houghton (WH) equation \cite{Wegner:1972ih} (see Ref.\,\cite{Clark:1992jr} 
for the detail form of $\beta_{\rm WH}$).
The right-handed side called the $\beta$-function is actually evaluated by the one-loop diagram
exactly. No higher loops do not contribute. 

This functional differential equation should be solved as the initial value problem, where the 
initial condition refers to the bare action,
\begin{align}
 S_{\rm eff} [\phi;0] = S_0[\phi] .
\end{align}
Solving the equation towards the infrared ($t \rightarrow \infty$) and we get the macro
effective action from which physical quantities such as the chiral condensate and the dynamical
quark mass are obtained.
The WH equation itself is exact, but it cannot be solved exactly, since it has infinite degrees of
freedom. 

Here we apply the WH equation to the NJL model. 
As an approximation, we restrict the full interaction space of the effective action 
$S_{\rm eff}[\psi,\bar\psi;t]$ into the subspace most relevant to D$\chi$SB.
We adopt the so-called local potential approximation where any quantum corrections to 
the derivative interactions are ignored. Furthermore, we set the local potential
as a function only of  the scalar fermion-bilinear field, $x=\bar\psi\psi$.
Then our effective action takes the following form,
\begin{align}
S_{\rm eff}[\psi,\bar\psi;t]&=
\int\!d^4x \left\{
\bar\psi\Slash{\partial}\psi - V(\bar\psi\psi;t)
\right\}.
\end{align}
The potential term $V(x;t)$, where $x$ denotes $\bar\psi\psi$ (integrated over space time), is called the fermion potential here, whose initial condition is set to be
\begin{align}
V(x;t=0)=\frac{G_0}{2}x^2
\end{align}
according to the NJL Lagrangian\,(\ref{eq:NJL}).

When evaluating the $\beta$-function we adopt an additional approximation of ignoring
the large-$N$ non-leading part in $\beta_{\rm WH}$.
Then, the NPRG equation for the fermion potential in the large-$N$ leading is given by 
the following partial differential equation,
\begin{align}
 \partial_t V(x;t)
=\frac{\Lambda^4}{4\pi^2}\log\left(1+\frac{1}{\Lambda^2}
(\partial_x V)^2\right)
\equiv - F(\partial_x V;t).
\label{eq:EqVw}
\end{align}
Hereafter we sometimes use a quick notation to save space like $\partial_t=\partial/\partial t$ {\sl etc}.
The renormalization scale $\Lambda$, the momentum cutoff, is defined by referring to 
the length of four-dimensional Euclidean momentum $p_\mu$,  that is, 
$\sum_{\mu=1}^{4} p_\mu^2 \leq \Lambda^2$. This is called the four-dimensional cutoff.
The right-hand side of equation is nothing but the simple trace-log formula of the Gaussian integral of the
shell modes and it corresponds to the one-loop corrections, though it is exact.
Note that the approximate NPRG equation above is known to be  equivalent to the mean field
calculation \cite{Aoki:1998gt,Aoki:1999dw,Aoki:1999dv}.

Now we introduce the mass function $M(x;t)$, the first derivative of the fermion potential, 
\begin{align}
M(x;t)=\partial_x V(x;t) .
\end{align}
The value of the mass function at the origin is the coefficient of mass operator $\bar\psi\psi$ 
in the effective action, and this is why we call it the mass function.
Note that it is still a function of $x$, that is, a function of operator, or more rigorously, a function of 
the bilinear Grassmannian variable $x$.

The chiral symmetry transformation defined in Eq.\,(\ref{eq:ChrlTrnsfmtn})  is represented by
the reflection,
\begin{align}
x  \longrightarrow -x.
\end{align}
The NPRG equation (\ref{eq:EqVw}) preserves this reflection symmetry and the initial condition
also respects it. Accordingly, the fermion potential at any {\sl time} $t$ satisfies the reflection
symmetry,
\begin{align}
V(-x;t)=V(x;t) ,
\end{align}
and therefore the mass function satisfies,
\begin{align}
M(-x;t)=-M(x;t) ,
\end{align}
and it is an odd function of $x$. 
Then the mass function must vanish at the origin,
\begin{align}
M(0;t)=0,
\end{align}
assuming it is well-defined at least.
This means there is no spontaneous mass generation and the phase is 
chiral symmetric.

Then what actually happens in case when D$\chi$SB and the spontaneous 
mass generation occurs?
We have to abandon well-definedness of the mass function at the origin, that is,  
the fermion potential $V$ must lose its differentiability at the origin. 
Then we can have the following situation,
\begin{align}
M(0+;t) = - M(0-;t) \ne 0,
\end{align}
that is, the mass function has a jump at the origin, still keeping its odd function property.
This is the spontaneous mass generation.
To explain clearly that this gives us the dynamically generated mass indeed, we need 
the notion of the effective potential, which will be introduced in Sec. 6.

Now we understand what we should deduce. 
At some finite $t_{\rm c}$, the fermion potential
becomes non-analytic at the origin, and it generates a finite gap in the mass function.
The gap continues to increase and finally gives the dynamical quark mass at the 
infrared. Just before $t_{\rm c}$, the second derivative of $V$ must become large and
increases towards infinite at $t_{\rm c}$. The second derivative of $V$ at the origin gives the four-fermion
interactions at the scale $t$, and this divergent behavior is nothing but the blowup solution
we find in the expanded NPRG equation in Eq.\,(\ref{eq:NJLbeta}) in case of $G_0>G_{\rm c}$
\cite{Aoki:2000wm,Braun:2011pp}.

See the plot (d) in Fig.\,\ref{fig:WKatMu0}, which is exactly what we like to realize
as a solution for the mass function. Before $t_{\rm c}$ the total function is a smooth
analytic function, whereas after $t_{\rm c}$ there is a finite jump at the origin which 
shows the spontaneous mass generation.
Such a {\sl solution}, however, is not allowed as a solution of the partial differential 
equation (PDE) in Eq.\,(\ref{eq:EqVw}), since the solution of the partial differential 
equation must satisfy the equation at every point. Therefore, the differentiability
at any point ($x,t$) is mandatory, while our expecting {\sl solution} must break it,  
though only at the origin.

To cure this situation, we have to modify the original equation so that 
differentiability of function $V$ might not be necessary.
This is the notion of weak solution, which is defined by the solution of the
weak equation, and it will be the subject of the next section.

\section{Weak solution and the Rankine-Hugoniot condition}

In this section, we first derive the partial differential equation for the mass function
and define the weak solution of it \cite{Sobolev:1938,Leray:1934}.
Then we construct the weak solution explicitly using the
standard method of the characteristics \cite{Evans:2010} in the next section.

We start with the following partial differential equation for $V(x;t)$,
\begin{align}
\partial_t V(x;t) = -F(\partial_x V, x;t).
\label{eq:EqVwX}
\end{align}
This is an extended system of the NJL-model  NPRG equation (\ref{eq:EqVw}),
where the $\beta$-function $F$ contains explicit dependence on $x$. 
The reason why we include this extension is that the gauge
theory analysis requires this additional $x$ dependence, and also the total
system becomes more symmetric and better interpreted in terms of the 
classical mechanics.

Differentiating the PDE\,(\ref{eq:EqVwX}) with respect to $x$, we obtain the following
PDE for the mass function,
\begin{align}
 \partial_t M(x;t)
& = -\partial_x F(M(x;t),x;t) \notag \\
& =-\frac{\partial F}{\partial M} \cdot
  \frac{\partial M}{\partial x}
 - \frac{\partial F}{\partial x} .
  \label{eq:StrgEq}
\end{align}
Here it should be noted that we have used somewhat subtle notations that 
two forms of the partial derivative of $F$
with respect to $x$ have different meanings as for treatment of variable $M$. 
This equation belongs to a class of PDE called the conservation law type since
it has a form of conservation law in two-dimensional space time.
This class contains
the famous Burgers' equation for non-linear wave without viscosity \cite{Burgers:1940}. 

We integrate the above equation convoluted with a smooth and bounded test function
$\varphi(x;t)$, and then by the partial integration we have 
\begin{align}
\int_{0}^{\infty}\!  dt   \int_{- \infty}^{\infty}  \! dx
\left(M \frac{\partial \varphi}{\partial t} 
+F\frac{\partial \varphi}{\partial x}\right )+ \int_{- \infty}^{\infty}&  dx  
~\left. M \,\varphi\right|_{t=0} =0.
\label{eq:WkEq}
\end{align}
Note that the above equation has no direct derivative of function $M(x;t)$.
Now we define the weak solution as follows;
the weak solution $M(x;t)$ satisfies Eq.\,(\ref{eq:WkEq}) 
for any smooth and bounded test function $\varphi$.
The weak solution $M(x;t)$ does not have to be differentiated, thus it can have
some non-differentiable point. 
Any strong solution satisfying the original PDE (\ref{eq:StrgEq}) at every point 
is in fact the weak solution. The inverse does not hold.
In this sense, the weak solution is a wider notion than the strong solution.
 
Then even in case when the strong solution does not have the global solution, 
the weak solution can be global and then we can calculate infrared physical 
quantities using the weak solution.
In general, the weak solution 
may not be unique. It also may depend on the initial condition. 
We will prove the weak solution of our NPRG equation is in fact unique and
global, and we can successfully obtain the infrared physics without any ambiguity. 

Here we state what is the weak solution without explicit proof
 (see {\sl e.g.} Ref.\,\cite{Evans:2010} for detailed argument). 
We suppose a weak solution has some jump singularities at some finite
number of points and otherwise it is smooth, that is, 
it is piecewisely differentiable.
Then almost everywhere other than the singularity, the weak solution 
must satisfy the strong equation (\ref{eq:StrgEq}) . 
This is the first property of the weak solution and thus it is not so different 
from the strong solution.

\begin{figure}
\centering
\includegraphics[width=0.4\hsize]{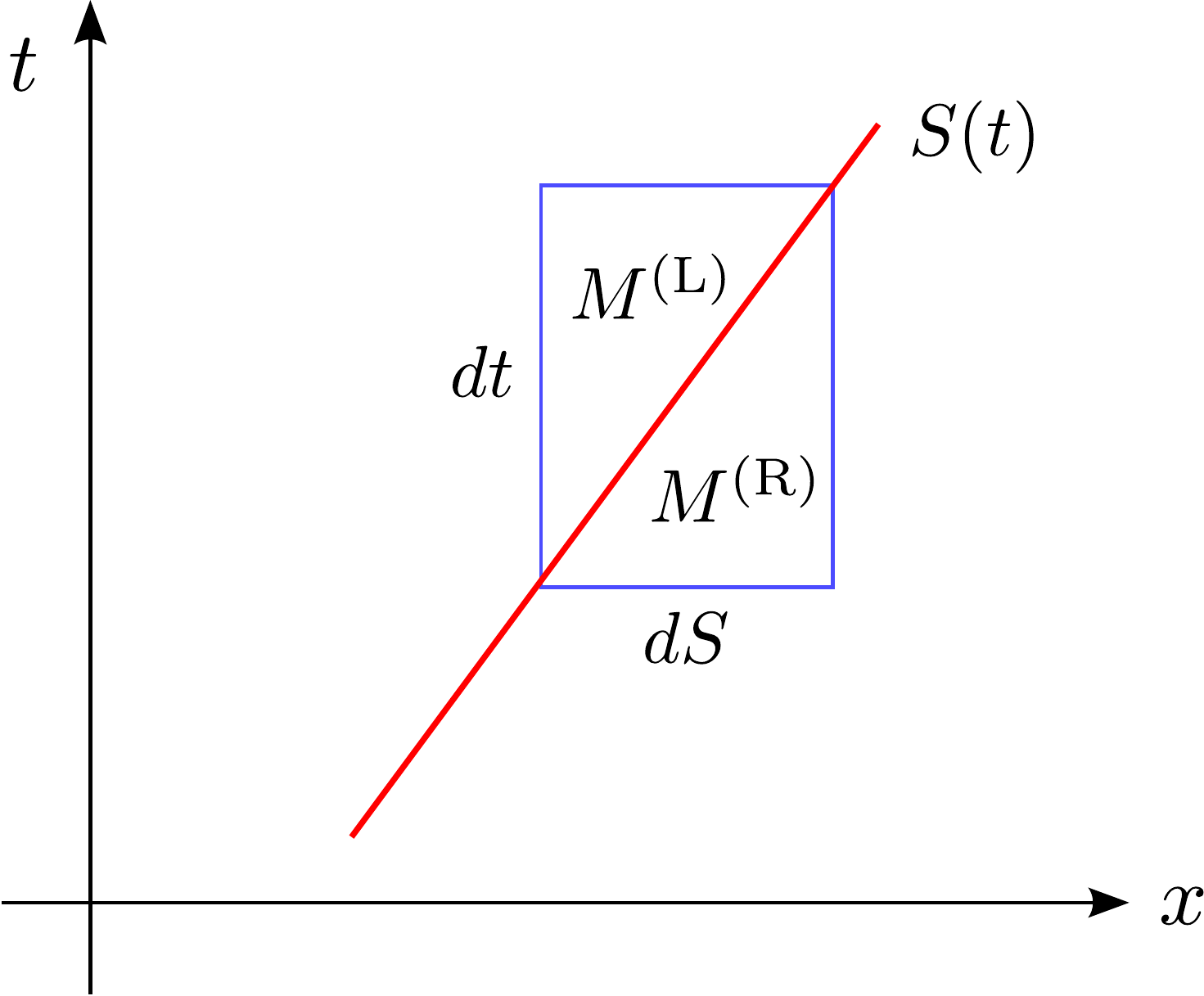}
\caption{Rankine-Hugoniot condition}
\label{fig:RHcondition}
\end{figure}

Second, what is required at the singularity?
As is shown in Fig.\,\ref{fig:RHcondition}, we look into the neighborhood 
of a singularity 
where there is a jump discontinuity on a curve $x=S(t)$, and values of $M$ at the
left-hand side and right-hand side of the singularity is denoted by $M_{\rm L}$ and 
$M_{\rm R}$ respectively.
Remember that the original equation (\ref{eq:StrgEq}) is regarded as the conservation law
where the {\sl charge} density is $M(x,t)$ and the {\sl current} is $F(M(x;t),t)$.
Consider a small boxed region $dS \times dt$ in Fig.\,\ref{fig:RHcondition}
and evaluate the total charge in the section of length $dS$.
The conservation law requires that the change of total charge must be balanced
with the difference between total flow-in and flow-out. We have,
\begin{align}
\left(M^{\rm (L)}-M^{\rm (R)}\right) dS(t) = 
\left[F(M^{\rm (L)})-F(M^{\rm (R)}) \right] dt.
\label{eq:RHcond}
\end{align}
This is nothing but the integrated form of the conservation law.

The above condition (\ref{eq:RHcond}) is called the Rankine-Hugoniot condition
\cite{Rankine:1870,Hugoniot:1887,Hugoniot:1889},
and all discontinuity must satisfy it. 
This is the second property which the weak solution must satisfy, and 
no more. 
This condition actually determines the 
equation of motion of the singularity position by the first order differential equation.
It needs one initial condition to determine the motion of the discontinuity point, 
and it is actually given by the spontaneous generation of the singularity as is seen later.

The Rankine-Hugoniot condition for the mass function $M$ is converted to the condition
for the fermion potential $V$. In Fig.\,\ref{fig:RHcondition}, we denote the fermion potential at
each side of the singularity by $V^{\rm (L)}$ and $V^{\rm (R)}$ respectively.
Then we evaluate its difference between bottom left point and top right point of the
$dt \times dS$ box as follows,
\begin{align}
dV^{\rm (L, R)}= 
\left.\frac{\partial V }{\partial x}\right|^{\rm (L,R)} dS
+ \left.\frac{\partial V }{\partial t}\right|^{\rm (L,R)} dt 
=M^{\rm (L,R)} dS -F(M^{\rm (L,R)}) dt.
\label{eq:dVw}
\end{align}
We have the following relation given by the RH condition (\ref{eq:RHcond}),
\begin{align}
dV^{\rm (L)} - dV^{\rm (R)} = (M^{\rm (L)}-M^{\rm (R)})dS 
- (F(M^{\rm (L)}) - F(M^{\rm (R)}))dt =0.
\label{eq:dVLR}
\end{align}
Therefore the fermion potential develops equally between left and right side of the
singularity. 
Taking account of the fact  
that there is no singularity in the beginning in case of our physical initial condition, 
that is, any singularity
starts in the mid of $t$-development, we conclude
\begin{align}
V^{\rm (L)} = V^{\rm (R)}.
\label{eq:RHV}
\end{align}
Thus the fermion potential function $V(x;t)$ is continuous everywhere as far as 
it is the weak solution of the PDE. This form of the RH condition helps us much
to obtain the weak solution in the next section.

\section{Method of characteristics}

In this section, we construct the weak solution using the standard textbook 
method of characteristics.
The characteristics is a curve on two-dimensional world ($x,t$) and is denoted by
$x=\bar{x}(t)$. Consider the mass function on this curve, $\bar M(t)=M(\bar{x};t)$,
and we calculate its derivative,
\begin{align}
\dfrac{d\bar M(t)}{dt}
&= \dfrac{\partial M(\bar x(t);t)}{\partial\bar x}  \dfrac{d\bar x(t)}{dt} 
+\dfrac{\partial M(\bar x(t);t)}{\partial t} \\
&=\dfrac{\partial M(\bar x(t);t)}{\partial\bar x}  \dfrac{d\bar x(t)}{dt} 
- \dfrac{\partial F(\bar M, \bar x(t);t)}{\partial \bar M} \dfrac{\partial M(\bar x(t);t)}{\partial \bar x}
- \dfrac{\partial F(\bar M, \bar x(t);t)}{\partial \bar x}.
\label{eq:barMdef}
\end{align}

We take the characteristics curve to satisfy the following differential equation,
\begin{align}
\dfrac{d\bar x(t)}{dt} & = 
\dfrac{\partial F(\bar M, \bar x;t)}{\partial \bar M}, 
\label{eq:barX}
\end{align}
then the mass function on it satisfies, 
\begin{align}
\dfrac{d\bar M(t)}{dt}
= -\dfrac{\partial F(\bar M, \bar x;t)}{\partial \bar x}.
\label{eq:barM}
\end{align}
This coupled pair of ordinary differential equations (\ref{eq:barX}) and (\ref{eq:barM})
are equivalent to the original two-dimensional partial differential equation
(\ref{eq:StrgEq}). 
The initial conditions for these equations are set by,
\begin{align}
\bar x(t=0)&=x_0, \\
\bar M(t=0)&=\left.\partial_x V(x;t)\right|_{x=x_0,t=0} = M_0,
\end{align}
where $x_0$ is a parameter discriminating characteristics and  $M_0$ is the
initial value of $M$ at  $x=x_0$. When explicitly indicating the initial value of the characteristic, we use
$\bar x(t;x_0)$.

\begin{figure}[h]
\centering
\begin{minipage}{0.28\hsize}
\centering
\includegraphics[scale=0.4]{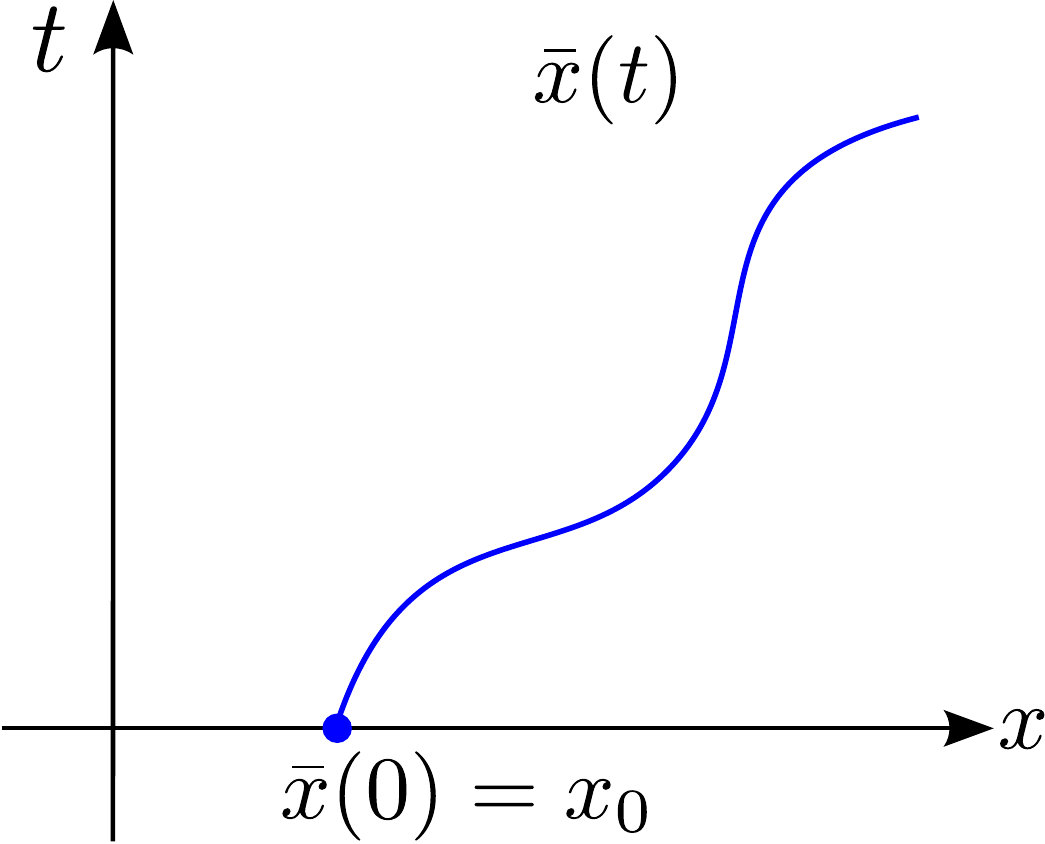}

{\small (a)}
\end{minipage}
\begin{minipage}{0.45\hsize}
\centering
\includegraphics[scale=0.4]{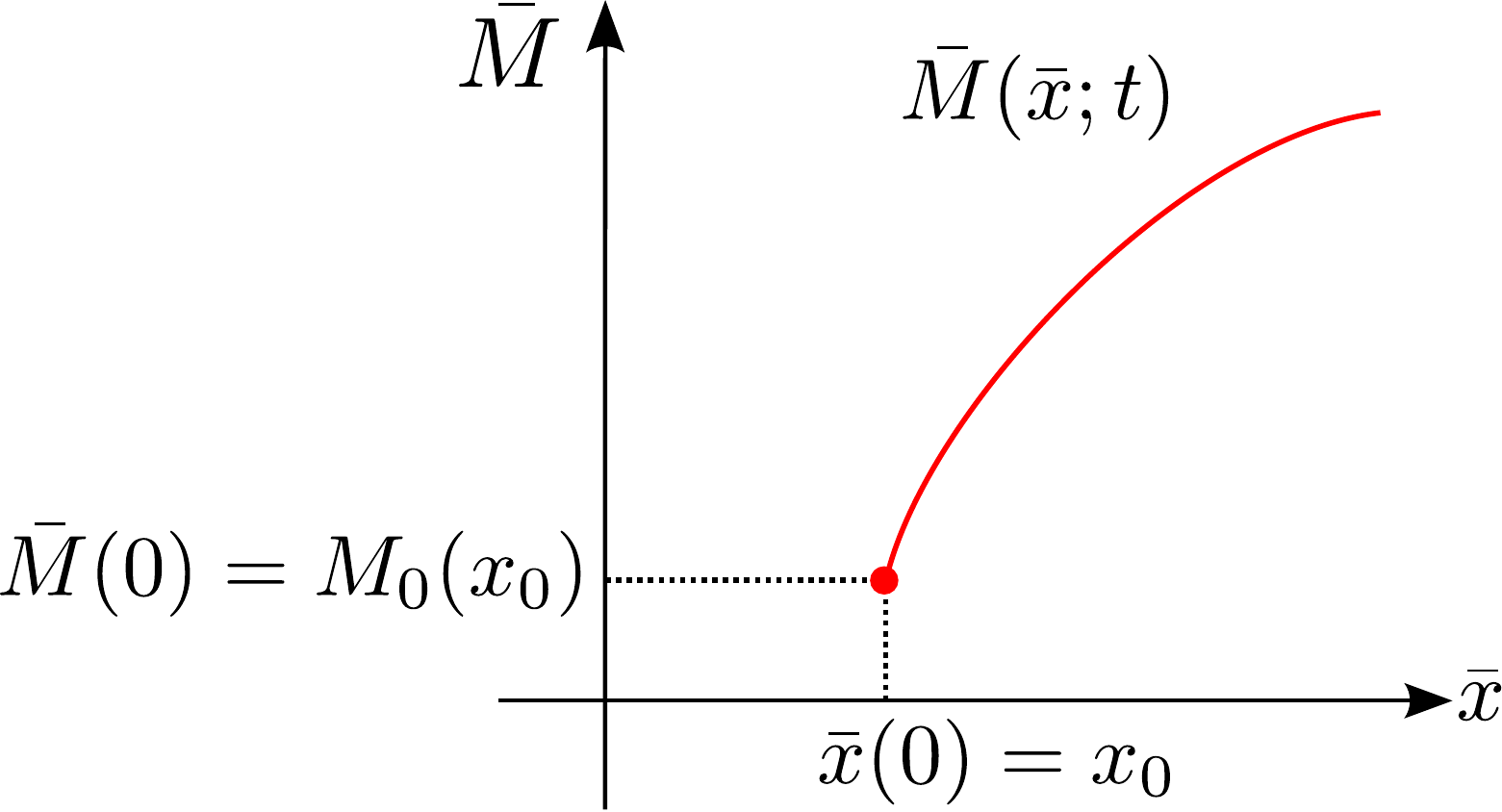}

\hspace{60pt}{\small  (b)}
\end{minipage}
\caption{Characteristics.}
\label{fig:char}
\end{figure}

The characteristics will give us the strong solution of PDE. 
Take a characteristics starting at $x_0$, we solve the coupled ODE, and then we get
the characteristics $\bar x(t; x_0)$ as shown in Fig.\,\ref{fig:char}\,(a), 
and the mass function on it, $\bar M(\bar x(t);t)$.

Here we change our viewpoint. 
The coupled ODE in equations (\ref{eq:barX}) and (\ref{eq:barM}) is regarded as the canonical
equation of motion of a kinematical system with the coordinate $\bar x$, the momentum $\bar M$ and
the {\sl time}-dependent Hamiltonian $F(\bar M, \bar x; t)$.
We plot two variables $\bar x(t)$ and $\bar M(t)$ in Fig.\,\ref{fig:char}\,(b)
and this is nothing but the phase space orbit of the Hamiltonian system.

The fermion potential $V$ also has its counterpart in mechanics.
Consider the fermion potential on the characteristics, 
$\bar V(t) = V(\bar x(t);t)$, and we calculate its derivative, 
\begin{align}
\frac{d \bar V(t)}{dt}
=\dfrac{\partial V(\bar x(t);t)}{\partial \bar x} \dfrac{d\bar x}{dt}
+ \dfrac{\partial V(\bar x(t);t)}{\partial t}
= \bar M \frac{d \bar x }{dt} -F(\bar M;t).
\label{eq:V_Lag}
\end{align}
The right-hand side is the {\sl Lagrangian} of this mechanical system.
Then the fermion potential is obtained by integration,
\begin{align}
V(\bar x(t);t) = V(x_0;t=0) + \int_0^t \left[\bar M \frac{d\bar x}{dt} -F(\bar M;t)\right]dt.
\label{eq:Vaction}
\end{align}
Thus this is the {\sl action} of the system as a function of the final time $t$ and 
the coordinate variable at the final time $\bar x(t)$, plus the initial value.
Then the original PDE in Eq.\,(\ref{eq:EqVwX}) is what is called the 
Hamilton-Jacobi equation in this kinematical system.

The starting point of this particle is controlled by the parameter $x_0$, and its initial 
momentum is the initial mass function $M(x_0;t=0)$.
Then we consider all orbits at once as shown in Fig.\,\ref{fig:stringmotion}, 
which is now the motion of string on the two-dimensional world $x \times M$.
The string never crosses to itself, since the equation of particle motion uniquely
determines the phase space orbit, thus no crossing of orbits is allowed.
The shape of the string at $t$ gives the mass function $M(x;t)$, the strong solution of original PDE.

\begin{figure}[h]
\centering
\includegraphics[scale=0.4]{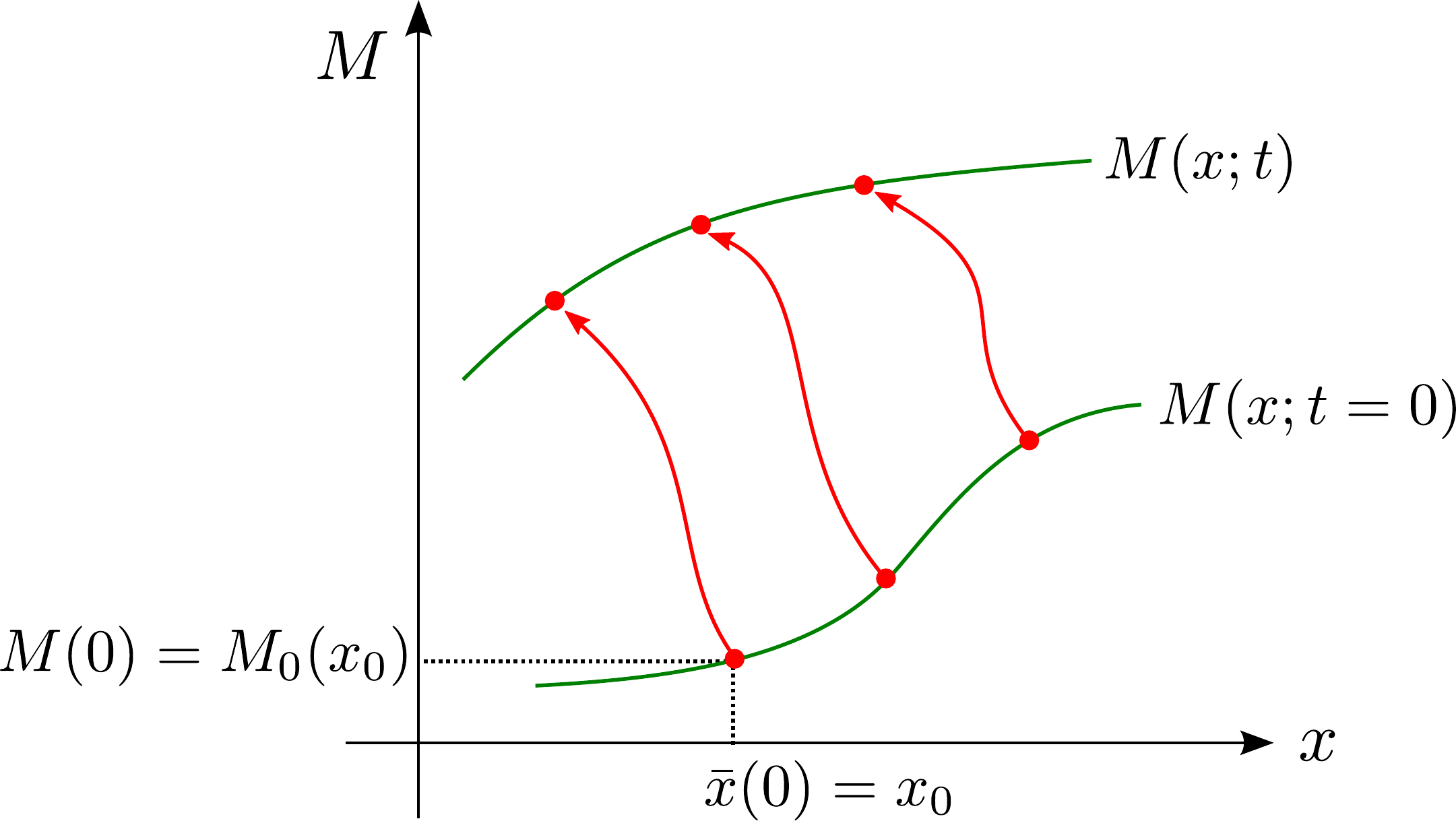}
\caption{Phase space orbit of particles and motion of string}
\end{figure}

There is nothing singular nor explosive blowup so far. 
The string motion gives the strong solution, locally. 
The point here is that it may not give the global solution as it is, when 
the string is folded and cannot define a function of $x$.
The string itself is parametrized by $x_0$, and $M(\bar x(t;x_0);t)$ is a unique
function of $x_0$. However when the characteristics are crossing to each
other, we have multi $x_0$ for single value of $\bar x(t)$. 
This situation is intuitively stated as that we have ``{\sl multi-valued function}''
 $M(x;t)$ at $t > t_{\rm c}$.

We note that the string motion here with the proper
initial condition has no singularity at all up to the infinite
time $t=\infty$. The string neither breaks up nor shrinks.
The string $M(\bar{x}(t;x_0);t)$ as a function of $x_0$ 
is always continuous. 
This fact will assure that the weak solution
we define now is unique and satisfies 
the so-called {\sl entropy condition}.
Normally the multi-folded structure is born as three-folded
leaves, and in the example in section 7, 
two three-folded structures meet together to become five-folded.

\begin{figure}[h]
\centering
\includegraphics[width=3.5in]{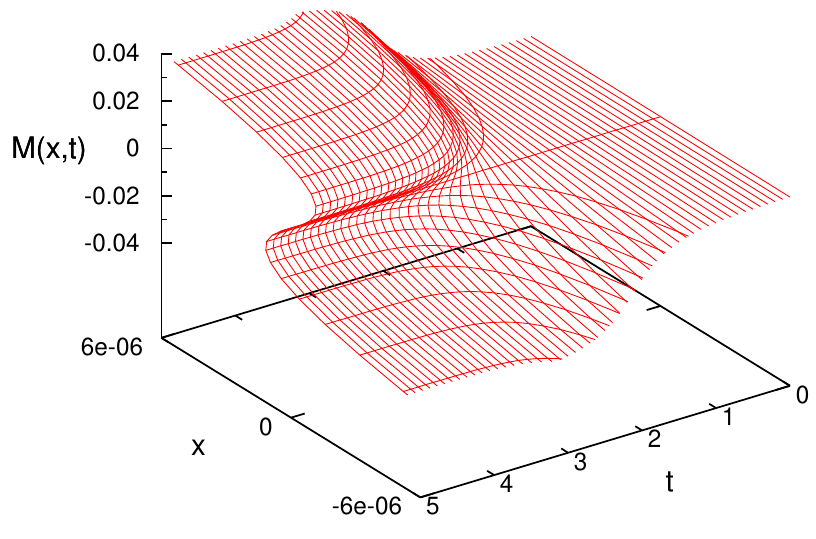}
\caption{World surface made of string motion.}
\label{fig:surface}
\end{figure}

Let us see a typical plot for the NJL model (super critical situation, $G_0=1.005 G_{\rm c}$) 
in Fig.\,\ref{fig:surface}, where the string motion and characteristics for 
$0<t<5$ is shown to make a surface.
The initial string is a straight line given by the initial mass function, $G_0 x$, and it develops to generate the self-folding structure. After a finite $t_{\rm c}$ the surface is  3-folded.
Simultaneously the characteristics pass over the origin.
All leaves of this surface are the strong solution at least locally.

Now we construct the weak solution by patch-working leaves to define
a single valued function $M(x;t)$ of $x$ at any time $t$.
After the patch-working, there appear discontinuity singularities.
How to cut and patch is determined by the RH condition.

\begin{figure}[h]
\centering
\begin{minipage}{0.45\hsize}
\centering
\includegraphics[scale=0.25]{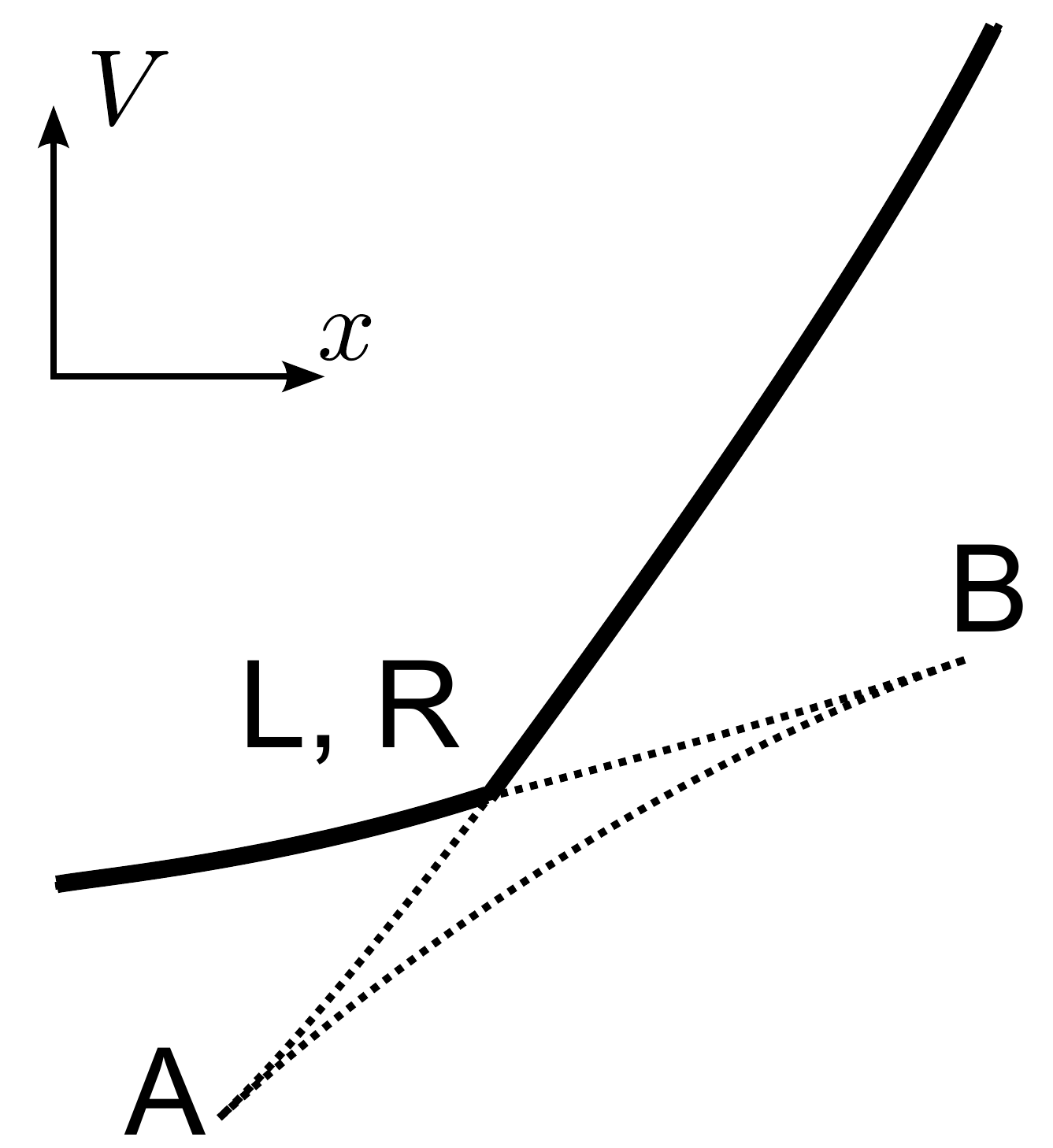}

{\small (a)}
\end{minipage}
\begin{minipage}{0.45\hsize}
\centering
\includegraphics[scale=0.25]{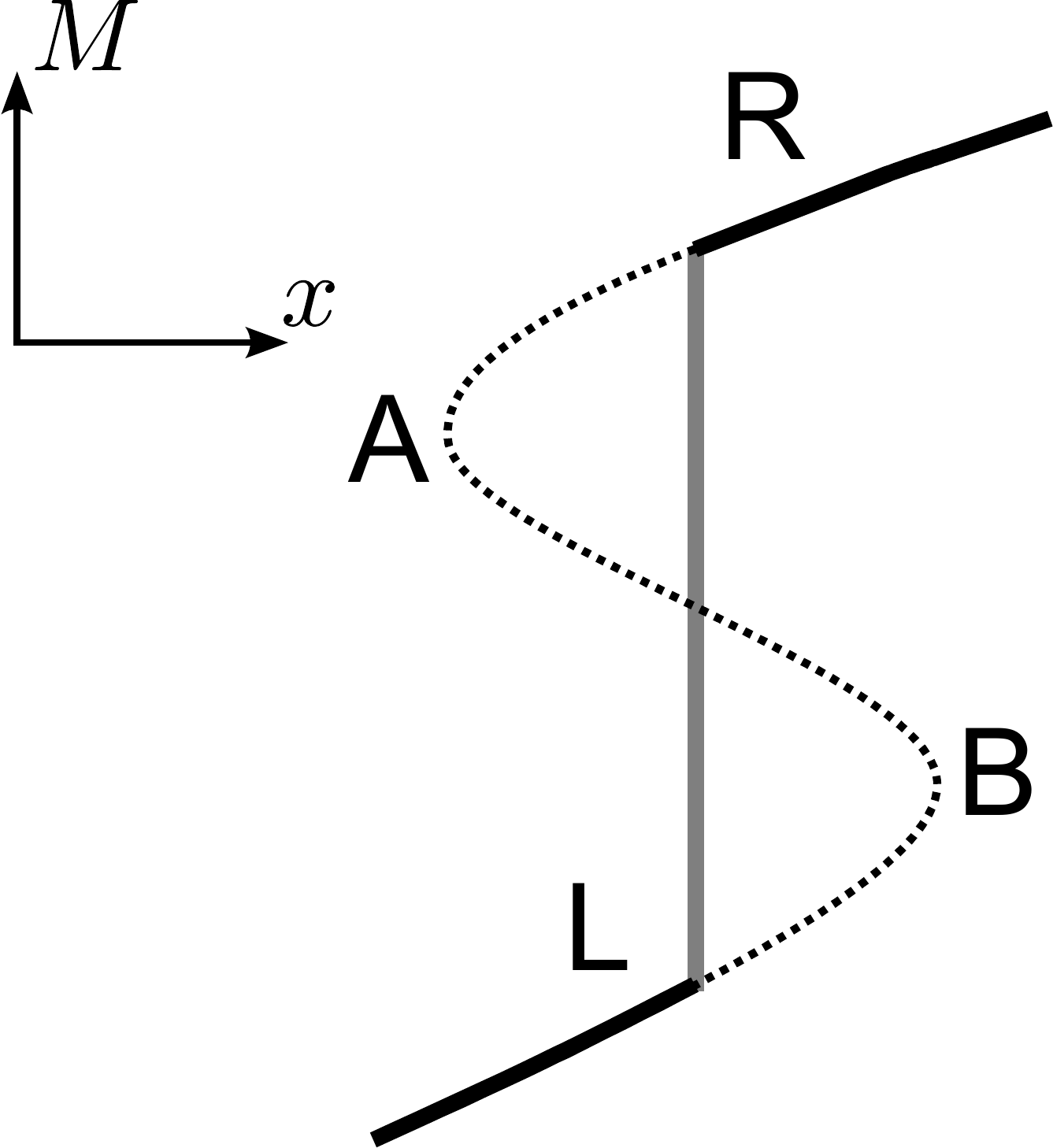}

{\small  (b)}
\end{minipage}
\caption{Geometrical interpretation of RH condition.}
\label{fig:eql_ar}
\end{figure}

We pick up a three folded sector and draw $V(x;t)$ and $M(x;t)$ as 
multi-valued functions of $x$ in Fig.\,\ref{fig:eql_ar} (a) and (b) 
respectively. 
The fermion potential has a characteristic shape of multi-folded structure, 
which is called {\sl swallowtail} in the mechanics 
or {\sl spinodal decomposition} in the thermodynamics.
The mass function $M(x;t)$ is a fixed $t$ section 
of the surface in Fig.\,\ref{fig:surface}.

The weak solution must satisfy the continuity of $V(x;t)$
according to the RH condition (\ref{eq:RHV}), and it is uniquely
determined to give the solid line as drawn 
in Fig.\,\ref{fig:eql_ar} (a).
Note that this selection can also be stated as that we take the maximum
possible branch among the candidates.

This weak solution gives a solution for
$M(x;t)$ as is drawn in Fig.\,\ref{fig:eql_ar} (b).
The points, A, B, L and R, in (a) and (b) are corresponding to each other, 
where L and R in (a) are the same.
The vertical line is the {\sl cut} we take for patch working 
and the mass function has a discontinuity from point L to R.
We integrate this multi-valued function along the string path, 
and we obtain
\begin{align}
\int_{\rm L}^{\rm R} Mdx = \int_{\rm L}^{\rm R} 
\frac{\partial V}{\partial x} dx
=V^{\rm (R)}-V^{\rm (A)}+V^{\rm (A)}-V^{\rm (B)}
+V^{\rm (B)}-V^{\rm (L)}
= V^{\rm (R)}-V^{\rm (L)}
=0,
\label{eq:RH_equal_area}
\end{align}
where we have used the RH condition in the form of Eq.\,(\ref{eq:RHV}).
This assures that the total (signed) area surrounded by the string and the cut
(vertical line) must vanish. In other words the two areas of left 
and right-hand side of the cut are equal to each other \cite{Whitham:1974}.

These geometrical representation of the RH condition is convenient to determine
the weak solution from the string motion diagram $M(x;t)$. 
When folding is generated, we just cut and patch so that 
we might keep the equal area rule.

We should stress here that the equal area rule is not a condition for the
weak solution but just a resultant feature referring to the strong solutions in the
backyard. These multi-leave strong solutions are not seen by the weak 
solution, and thus it is impossible to refer to the backyard solutions in 
order to get the next time step. The target solution in the renormalization
group equation is the effective interactions at a scale $t$ and there is 
absolutely no information for the backyard strong solutions.
In fact, the original RH condition (\ref{eq:RHcond}) is equivalent to the conservation of 
the difference $V^{\rm (R)}-V^{\rm (L)}$ in Eq.(\ref{eq:dVLR}).
After adding the fact that any singularity is generated 
in the mid of $t$ development, we have the continuity 
$V^{\rm (R)}=V^{\rm (L)}$ and consequently the equal area rule.
Original condition (\ref{eq:RHcond}) or (\ref{eq:dVLR}) 
is enough to determine the weak solution uniquely.
Therefore we do not have to do any access to the backyard solutions to
solve the renormalization group equation. The renormalization group equation
determines its weak solution by itself, without recourse to the multi-leave solutions.

The above equal area rule may remind the reader of the Maxwell's rule in 
treating the coexisting situation, for example, the liquid and the gas, where the
vertical line in Fig.\,\ref{fig:eql_ar} (b) looks like the coexisting isothermal line 
in the $p(=x)$-$V(=M)$ phase diagram. However the Maxwell's rule has to refer to 
the backyard thermodynamic functions which are usually not well established
in the equilibrium thermodynamics. In this sense the Maxwell's rule is to be
classified as a phenomenological rule. Nevertheless we may assign such
jump line between L and R the coexisting states of L and R states with the
bulk domain structure. We do not discuss this viewpoint more since it is
out of the scope of this paper.

As is proved in Eq.\,(\ref{eq:Vaction}), the fermion potential is the {\sl action} 
up to the initial constant in the mechanics analogy, 
and in fact the selection of leaves satisfying the RH condition  
will be done so that the {\sl maximum action} 
principle holds, which is directly related to the minimum free energy principle
as is seen later.

\section{Weak solution of the NJL model}

\begin{figure}[ht]
\centering
\includegraphics[scale=0.4]{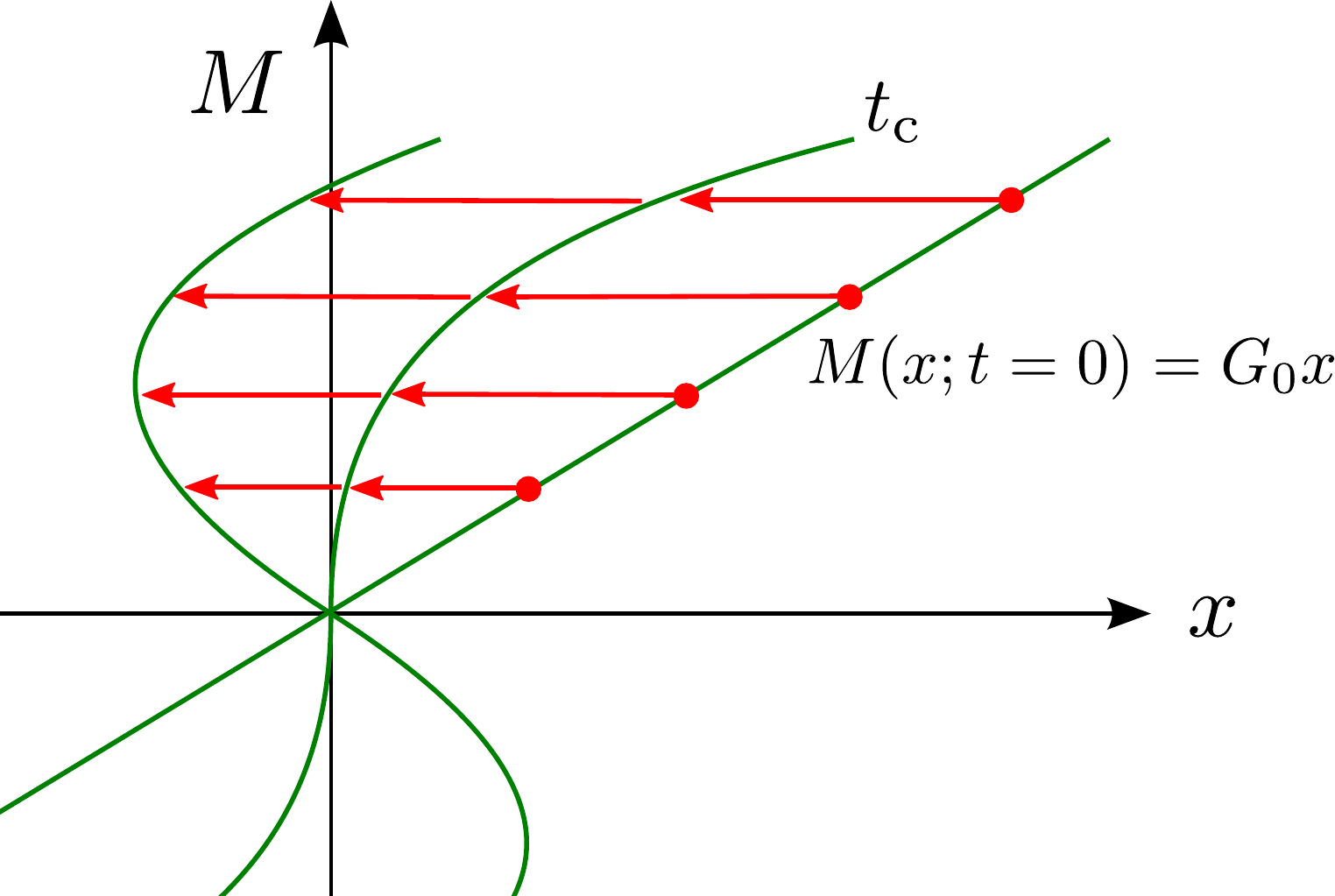}
\caption{String motion.}
\label{fig:stringmotion}
\end{figure}

We construct the weak solution for the NJL model explicitly.
The {\sl Hamiltonian} $F$ has no explicit dependence on $x$, that is, 
the system is {\sl translationally} invariant,
\begin{align}
F(M,x;t) = - \frac{\Lambda^4}{4\pi^2}\log\left(1+\frac{M^2}{\Lambda^2}
\right).
\label{eq:NJL_Fterm}
\end{align}
Therefore the {\sl momentum} $M$ is conserved in the {\sl time} development
and the characteristics $\bar x(t)$ are the contour of $M$. 
Then the string motion is driven by horizontal move of each point as shown in
Fig.\,\ref{fig:stringmotion}.
The initial position of the string is given by the bare four-fermi interactions,
\begin{align}
M(x;0) = \partial_x V(x;0) = G_0 x,
\end{align}
which is a straight line.

The {\sl velocity} of each point of string is given by, 
\begin{align}
\dfrac{d\bar x(t)}{dt}= \dfrac{\partial F(\bar M,x;t) }{\partial \bar M} 
=  - \dfrac{\Lambda^4}{4\pi^2} \dfrac{2 M }{\Lambda^2 + M^2},
\label{eq:NJL_char}
\end{align}
and it has an opposite sign against the momentum $M$. 
The velocity is suppressed for large $M$ region and it will vanish for large $t$ (small $\Lambda$).
The schematic motion of 
string is drawn in Fig.\,\ref{fig:stringmotion} in case of super critical phase, where
the string starts being folded after a finite $t_c$. 

Now we solve the characteristics by integrating Eq.\,(\ref{eq:NJL_char}) with 
the initial conditions,
\begin{align}
\bar x(0;x_0)=x_0\ , \ \ \bar M(x(0);0) = G_0 x_0 =M_0,  
\end{align}
and we get,
\begin{align}
\bar x(t;x_0) & =x_0 + \int^t_0 \frac{d\bar x}{dt} dt 
=x_0 -\frac{2M_0\Lambda^4_0}{4\pi^2}
  \int^t_0 \frac{{\rm e}^{-4t}}{\Lambda_0^2 {\rm e}^{-2t} +M_0^2} dt
\notag \\
&= x_0
  +\frac{\Lambda_0^2 M_0}{4\pi^2}
   \left[
    {\rm e}^{-2t} - 1
    -\frac{M_0^2}{\Lambda_0^2}\log
     \frac{\Lambda_0^2{\rm e}^{-2t} +M_0^2}{\Lambda_0^2+M_0^2} 
   \right] \notag\\
&\equiv C_t(x_0).
\end{align}
If $t$ is enough small, the function $C_t$ is a monotonic function and then
we can express the initial value $x_0$ by a function of the final value 
$x(t)$ using the inverse of function $C_t$,
\begin{align}
x_0 = C^{-1}_t(x).
\label{eq:inverse_C}
\end{align}
Finally we obtain the solution by using this inverse function as follows,
\begin{align}
M(x;t)= M(C^{-1}_t(x);0)
      = G_0 C_t^{-1}(x).
\label{eq:NJL_strong_sol}
\end{align}

The solution in Eq.\,(\ref{eq:NJL_strong_sol}) is the strong solution 
as far as  the inverse function $C^{-1}_t$ can be defined.
This situation will be broken at finite $t$ for strong enough coupling 
constant $G_0$. The monotonicity of function $C_t$ breaks at the 
origin first. We evaluate the derivative at the origin,
\begin{align}
 \left. \frac{dC_t (x_0)}{dx_0}\right|_{x_0=0}
=\left. \frac{\partial \bar x(t;x_0)}{\partial x_0}
\right|_{x_0=0}
=1+\frac{\Lambda_0^2G_0}{4\pi^2} \left({\rm e}^{-2t}-1\right).
\end{align}
This becomes negative at a finite $t_{\rm c}$ ,
\begin{align}
t_{\rm c}=\frac{1}{2} \log\frac{G_0\Lambda_0^2}{G_0\Lambda_0^2-4\pi^2},
\end{align}
for strong bare coupling constant $G_0 > G_c$,
\begin{align}
G_{\rm c}\Lambda_0^2=4\pi^2\ .
\end{align}
This result is equal to the argument in Sec.\,1 using the blowup solution.

Finally we calculate the fermion potential exactly by integrating Eq.\,(\ref{eq:V_Lag}),
\begin{align}
V(x;t)&= V(x_0;0) +\int_0^t\frac{dV(\bar x(t);t)}{dt} dt \notag \\
 &= \dfrac{1}{2}G_0 x_0^2 +\frac{3}{4} M_0 \left(x - x_0 \right)
   -\frac{\Lambda_0^4}{16\pi^2}
   \left[
     {\rm e}^{-4t} \log\left(1+\frac{M_0^2}{\Lambda_0^2} {\rm e}^{2t}\right)
     -\log\left(1+\frac{M_0^2}{\Lambda_0^2}\right)
    \right]\notag\\
 &= \dfrac{1}{2}G_0 x_0^2 +\frac{\Lambda_0^2}{16\pi^2}
    \Bigg\{
    \left[
     3M_0^2 \left(
     {\rm e}^{-2t} -1
     -\frac{M_0^2}{\Lambda_0^2}
      \log\frac{\Lambda_0^2 {\rm e}^{-2t}+M_0^2}{\Lambda_0^2+M_0^2}
     \right)
    \right] \notag \\
    &\qquad
    -\Lambda_0^2\left[
     {\rm e}^{-4t} \log\left(1+\frac{M_0^2}{\Lambda_0^2} {\rm e}^{2t}\right)
     -\log\left(1+\frac{M_0^2}{\Lambda_0^2}\right)
    \right]
    \Bigg\}\ ,
\end{align}
where $x_0$ and $M_0=G_0 x_0$ should be understood as a function of $x$  
through the inverse function $C_t^{-1}$ in Eq.\,(\ref{eq:inverse_C}).
The fermion potential is also multi-valued where the inverse function $C_t^{-1}$
is multi-valued. 

\begin{figure}[h]
\centering
\begin{minipage}{0.4\hsize}
\centering
\includegraphics[width=\hsize]{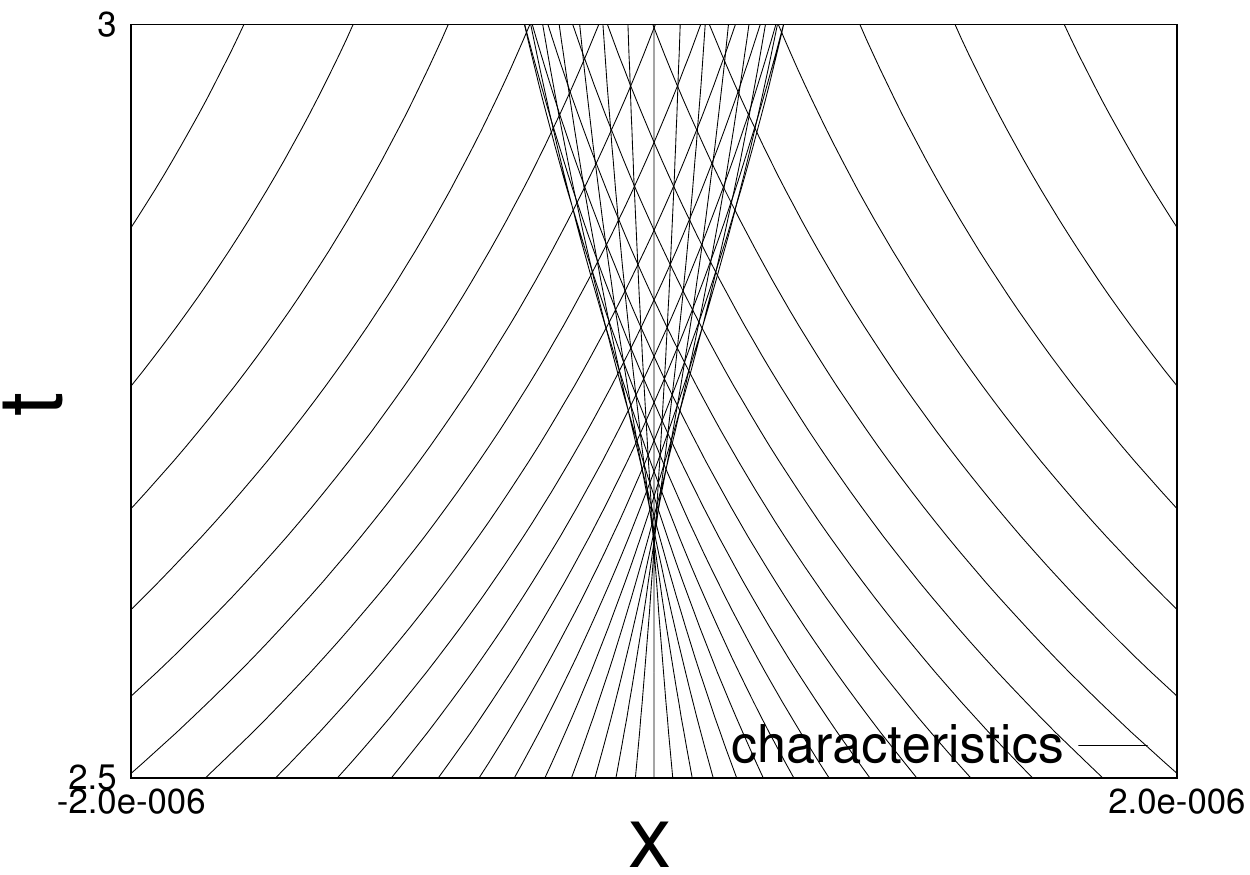}
\\ \hspace{8pt}{\small (a)}
\end{minipage}
\hspace{3pt}
\begin{minipage}{0.4\hsize}
\centering
\includegraphics[width=\hsize]{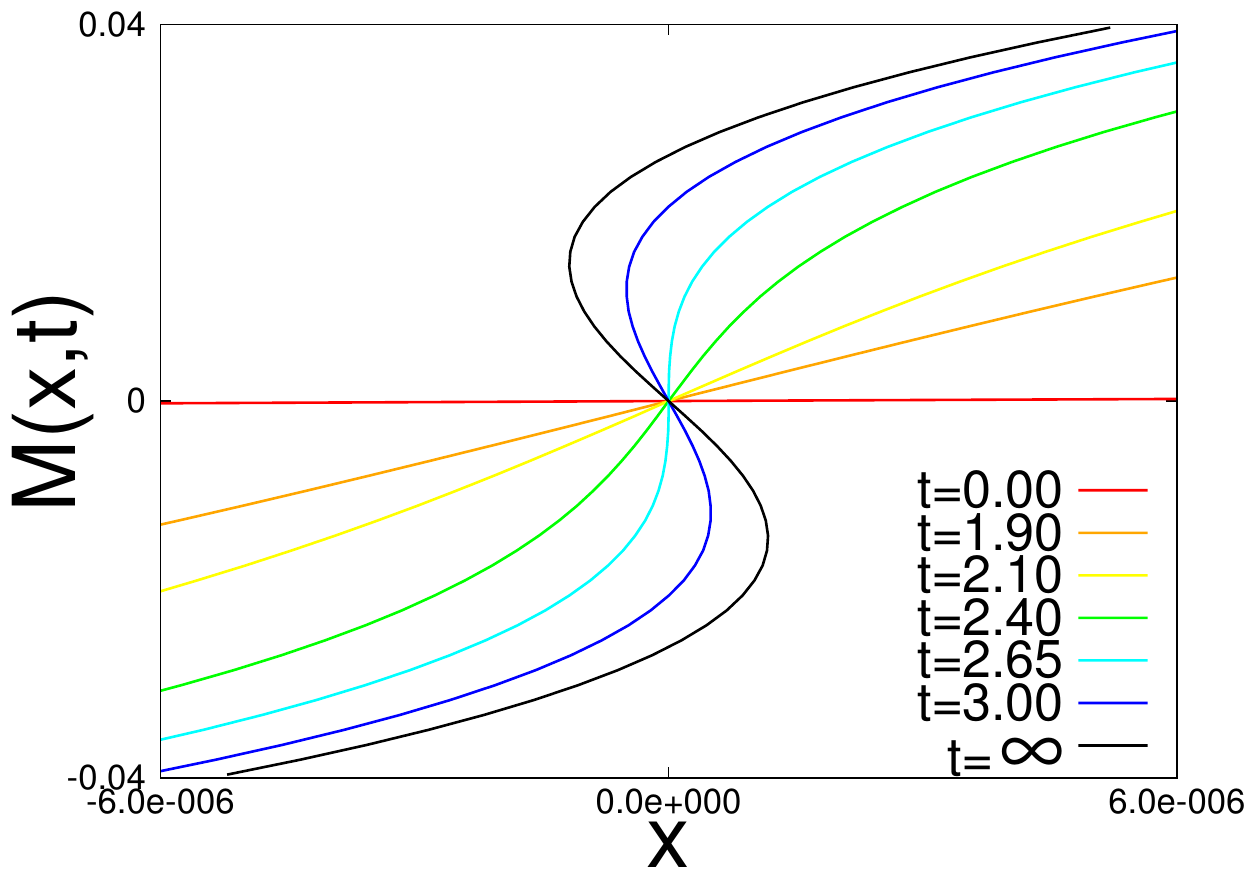}
\\ \hspace{14pt} {\small (b)}
\end{minipage}
\vskip 10pt
\begin{minipage}{0.4\hsize}
\centering
\includegraphics[width=\hsize]{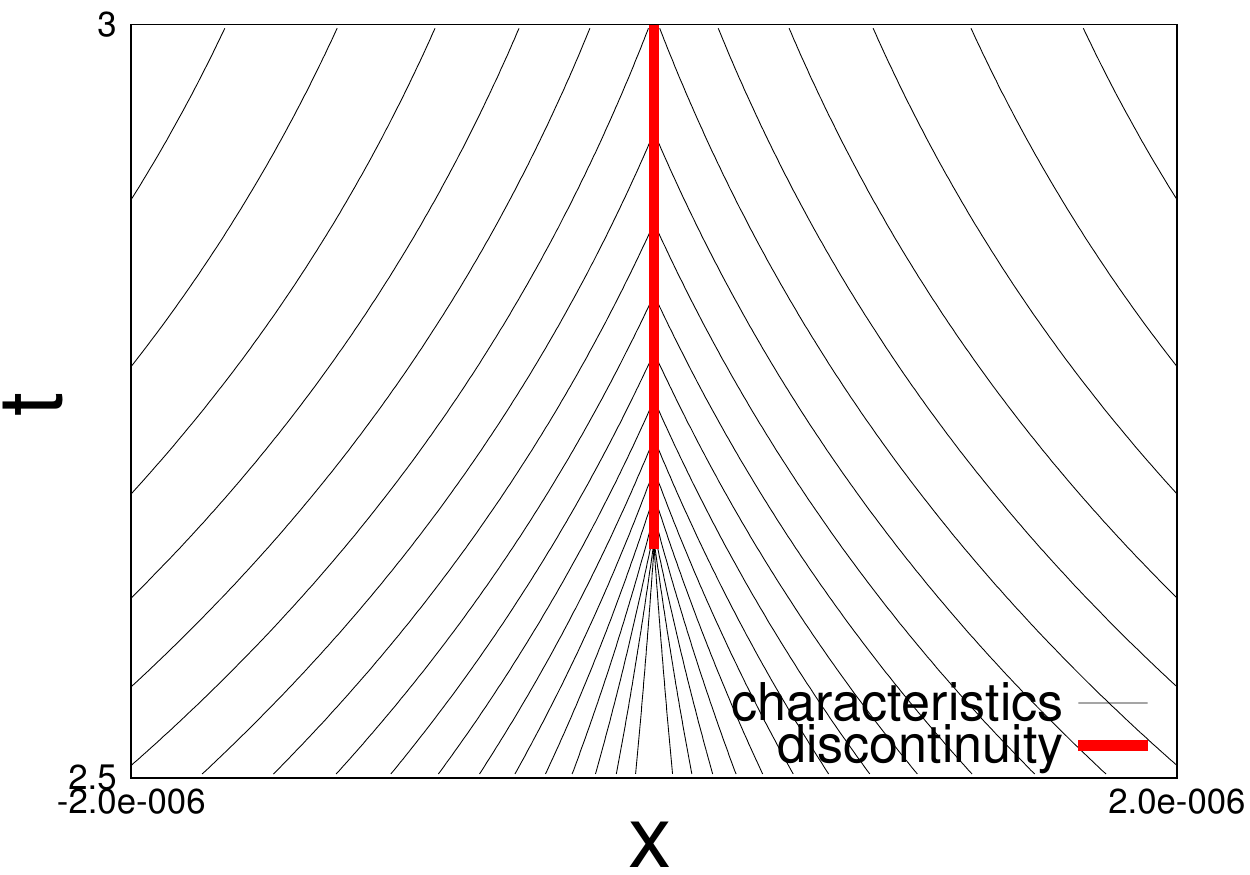}
\\ \hspace{8pt} {\small (c)}
\end{minipage}
\hspace{3pt}
\begin{minipage}{0.4\hsize}
\centering
\includegraphics[width=\hsize]{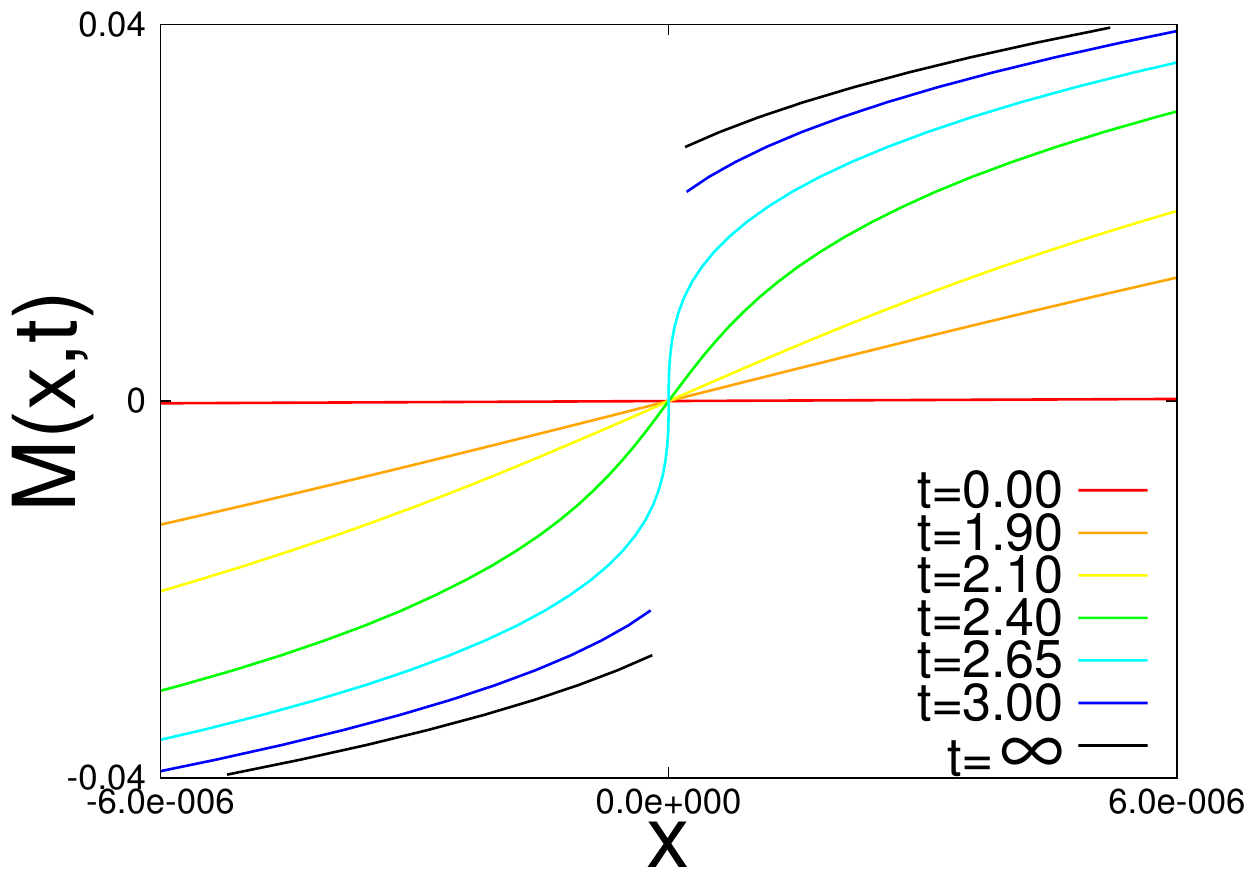}
\\ \hspace{14pt} {\small (d)}
\end{minipage}
\caption{(a) Characteristics. (b) String motion, i.e., set of local strong solutions of mass function. 
(c) Characteristics selected by the RH condition and discontinuity at the red line. 
 (d) Weak solution of mass function. Discontinuity at the origin to satisfy RH condition.}
\label{fig:WKatMu0}
\end{figure}

We show the numerical plots of these results in Fig.\,\ref{fig:WKatMu0}
 in case of the super critical interaction, $G_0=1.005 G_{\rm c}$ and $t_{\rm c}=2.65$, 
and explain how to do the patch work to construct the weak solution. 
Hereafter all the dimensional variables are rescaled by $\Lambda_0$ to be dimensionless.
The characteristics are plotted in Fig.\,\ref{fig:WKatMu0}(a), and we find  they will start
crossing to each other after $t>t_{\rm c}$. The over populated region enlarges with 
3-folded structure.
The {\sl string} motion is drawn in Fig.\,\ref{fig:WKatMu0}(b), where we see it starts
folding after $t_{\rm c}$. 

Then we have to cut and patch after $t_{\rm c}$.
First, note that the original chiral symmetry of the system assures the odd function property of
the mass function $M(-x;t) = -M(x;t)$ even after folding. 
The RH condition of the form in Eq.\,(\ref{eq:RH_equal_area}) tells us 
that the cut out area should have the same area always.
Then the only way to satisfy RH condition is to put the discontinuity at the origin
and adopt the up and down leaves, without using the mid leaf.
Thus, we obtained the weak solution, which are drawn in Fig.\,\ref{fig:WKatMu0}(c)/(d).
This weak solution is uniquely obtained for any initial four-fermi coupling constant 
and it is always global up to the infrared limit $t\rightarrow\infty$.

From the conservation law point of view, the symmetry breaking and the RH solution after that are understood as follows. 
The {\sl current} $F$ is  always negative and it is larger for larger $|M|$. 
Then the charge $M$ accumulates towards the origin with increasing slope $\partial_x M|_0$ since the current at the origin is vanishing. At $t_{\rm c}$, 
the slope becomes infinite, the singularity appears and we start defining discontiuous solution.
Now the charge flows in to the backyard and flows out from the backyard at the origin, ignoring the vanishing current condition there.
This is the RH solution.

We answer here for the possible question for the necessity of the weak solution.
If we agree that there is a singularity in $M$ in fact, but only at the origin, then according to 
the total symmetry $M(-x)=M(x)$, we may solve the system only for the half space
$(0,\infty)$. This is true. We demonstrated that the numerical analysis of the PDE
in $(0,\infty)$ with the free end boundary condition at $0+$ 
could work actually and $M(0+)$ is obtained to grow up after $t>t_{\rm c}$
\cite{Aoki:2012mj}.
However this way of getting the dynamical mass is  really 
tedious and its reliability is not proved.
Moreover, most important is that such trick does not work  for the first order
phase transition case in Sec.7, where the singularity appears off the origin,
pairwisely, and then they move. 
In such case the weak solution is the only method to reliably calculate the 
physical quantities.

\section{Bare mass and the Legendre effective potential}

In the previous section, we have obtained the D$\chi$SB weak solution of the mass function $M(x;t)$ 
which has a finite jump at the origin.
In this section we argue how to calculate the chiral order parameters,
the dynamical mass of quark and the chiral condensate, in the infrared limit.

The physical mass of quark should be defined by $M(0;\infty)$ and
it is actually not well-defined even by the weak solution itself.
This is a common issue of the spontaneous symmetry breaking and we take a standard method of 
adding  the bare mass term $m_0\bar\psi\psi$
explicitly breaking the chiral symmetry to the bare action and investigate what happens.

The addition of bare mass term does not modify our system of NPRG at all. It just changes the
initial condition of the fermion potential as
\begin{align}
V(x;0) = \dfrac{G_0}{2} x^2 + m_0 x,
\end{align}
and the initial mass function takes the following form,
\begin{align}
M(x;0) = G_0 x + m_0.
\end{align}
As is stressed in Eq.\,(\ref{eq:NJL_Fterm}), the NJL system is translationally invariant with respect to
$x$. We  shift the origin of $x$ coordinate as,
\begin{align}
x \longrightarrow x - m_0/G_0\ ,
\end{align}
and completely the same calculation holds for the massless case in the previous section.
Therefore the weak solution with bare mass is obtained from that of massless case by
shifting  $x$ argument as follows,
\begin{align}
 M(x;t,;m_0)= M(x+m_0/G_0;t,;m_0=0).
\end{align}
As for the fermion potential, we have to take account of the global shift of the bare potential
in addition to the shift of $x$,
\begin{align}
V(x;t;m_0) = V(x+m_0/G_0;t;m_0=0) - \dfrac{m_0^2}{2G_0}
\label{eq:V_shift}
\end{align}
This looks miraculous considering that the RGE system with bare mass is completely 
different from massless case, all operators are coupled together.

\begin{figure}[h]
\centering
\includegraphics[width=0.5\hsize]{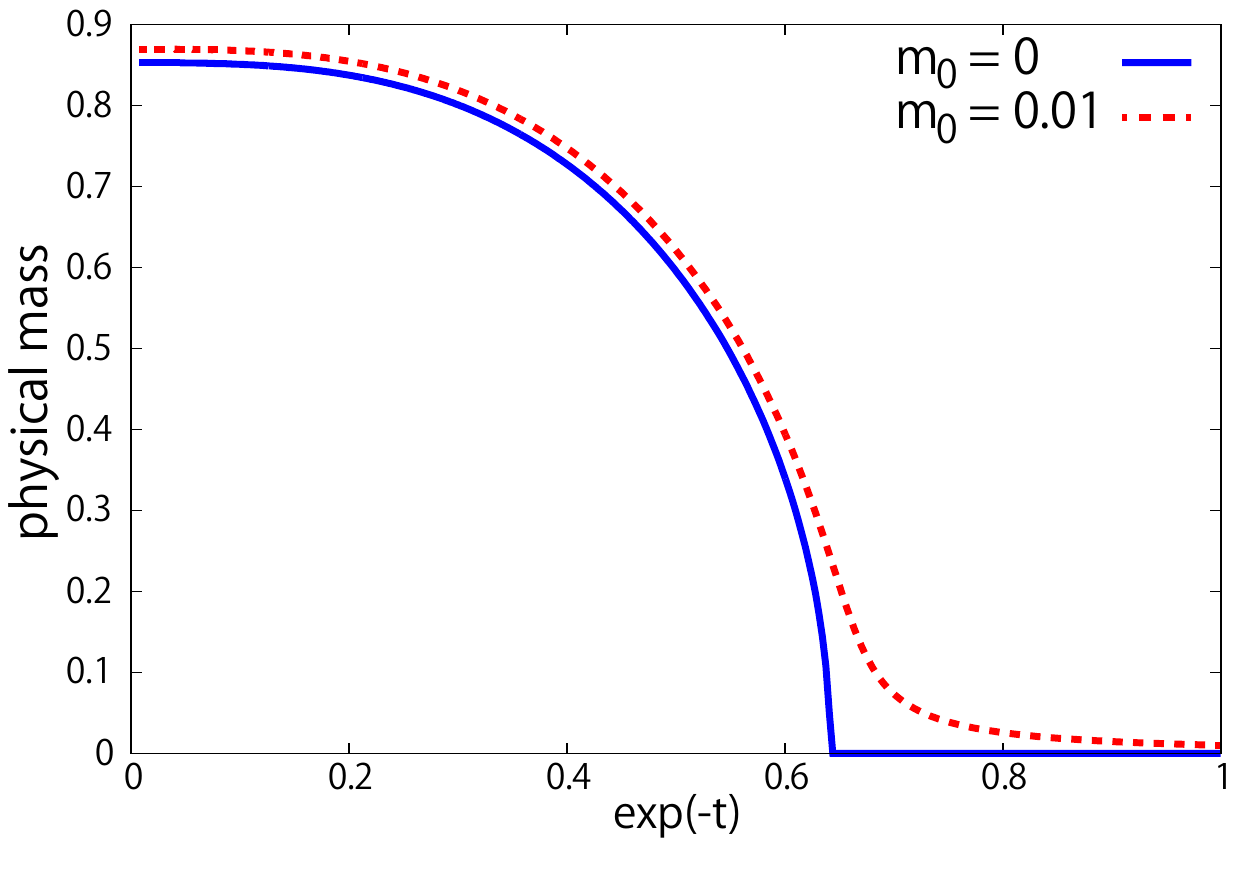}
\caption{RG evolution of the physical masses in $m_0\rightarrow 0$ and $m_0=0$. 
The NPRG equation given by Eq.\,(\ref{eq:FatMuNeq0}) ($\mu=0$, $G=1.7G_{\rm c}$)
 is used.}
\label{fig:2nd_physmss}
\end{figure}

Now the mass function at the origin is well-defined  and we can define the physical mass by
the following chiral limit,
\begin{align}
M_{\rm phys}(t)= \lim_{m_0\rightarrow +0} \left.M(x;t,;m_0)\right|_{x=0}.
\end{align}
Fig.\,\ref{fig:2nd_physmss} shows the RG evolution of the physical mass 
for $m_0\ne 0$ and its chiral limit.
The physical mass in the chiral limit shows the second order phase transition 
due to the singular behavior of the mass function at the origin, 
while the physical mass at $m_0\neq 0$ shows the cross over behavior.
 
The reader may consider that the weak-solution method is not necessary if we consider $m_0\neq 0$
from the very beginning.
This statement is absolutely wrong.
First of all, note that the mass function is a function of operator $x$ totally describing the effective interactions.
Even with the bare mass included, the mass function still has completely the same singularity. 
Singularity does not disappear, and it is just moved away from the origin.
Actually, Ref.\,\cite{Aoki:2012mj} shows that the Taylor expansion method to solve 
the PDE\,(\ref{eq:EqVw}) does not work with the small bare mass.  

Second, if we are interested only in the origin of the mass function, then there seems to be no 
singularity as just drawn in Fig.\,\ref{fig:2nd_physmss}. However this is the case of the second order
phase transition. Such disappearance of singularity does not hold for the first order phase
transition, which will be explicitly argued in the next section.
The difference here is that the second order phase transition is directly related to the symmetry 
breakdown and therefore is sensitively affected by the explicit breaking term,  
while the first order phase transition is not related at all to the symmetry breaking
and therefore is not altered much by the explicit breaking term. 
These points will be clear in the next section.

In order to evaluate the chiral condensate and the Legendre effective potential for it, 
we first introduce the source term $j\bar\psi\psi$ in the bare action.
Then the initial condition of the fermion potential is 
\begin{align}
V(x;t=0;j)&= m_0 x + \frac{G_0}{2} x^2 +jx.
\end{align}
The free energy as a function of $j$ is given by the value of $V$ at the origin, 
\begin{align}
W(j;t)=V(x=0;t;j),
\end{align}
and the chiral condensate is defined by,
\begin{align}
\phi(j;t)\equiv \langle \bar{\psi}\psi\rangle_j
=\frac{\partial W(j;t)}{\partial j}.
\label{eq:defofCC}
\end{align}
Note that these quantities are all defined at renormalization scale $t$.

We define the Legendre effective potential of the chiral condensate as follows,
\begin{align}
V_{\rm L} (\phi;t)= j\phi(j;t) -W(j;t),
\label{eq:Legendre}
\end{align}
where $j$ in the right-hand side is considered as a function of $\phi$ through
the inverse of Eq.\,(\ref{eq:defofCC}).
We have the standard relation,
\begin{align}
\dfrac{\partial V_{\rm L}}{\partial \phi}=j.
\end{align}

\begin{figure}[h]
\centering
 \begin{minipage}{0.4\hsize}
  \centering  
  \includegraphics[scale=0.4]{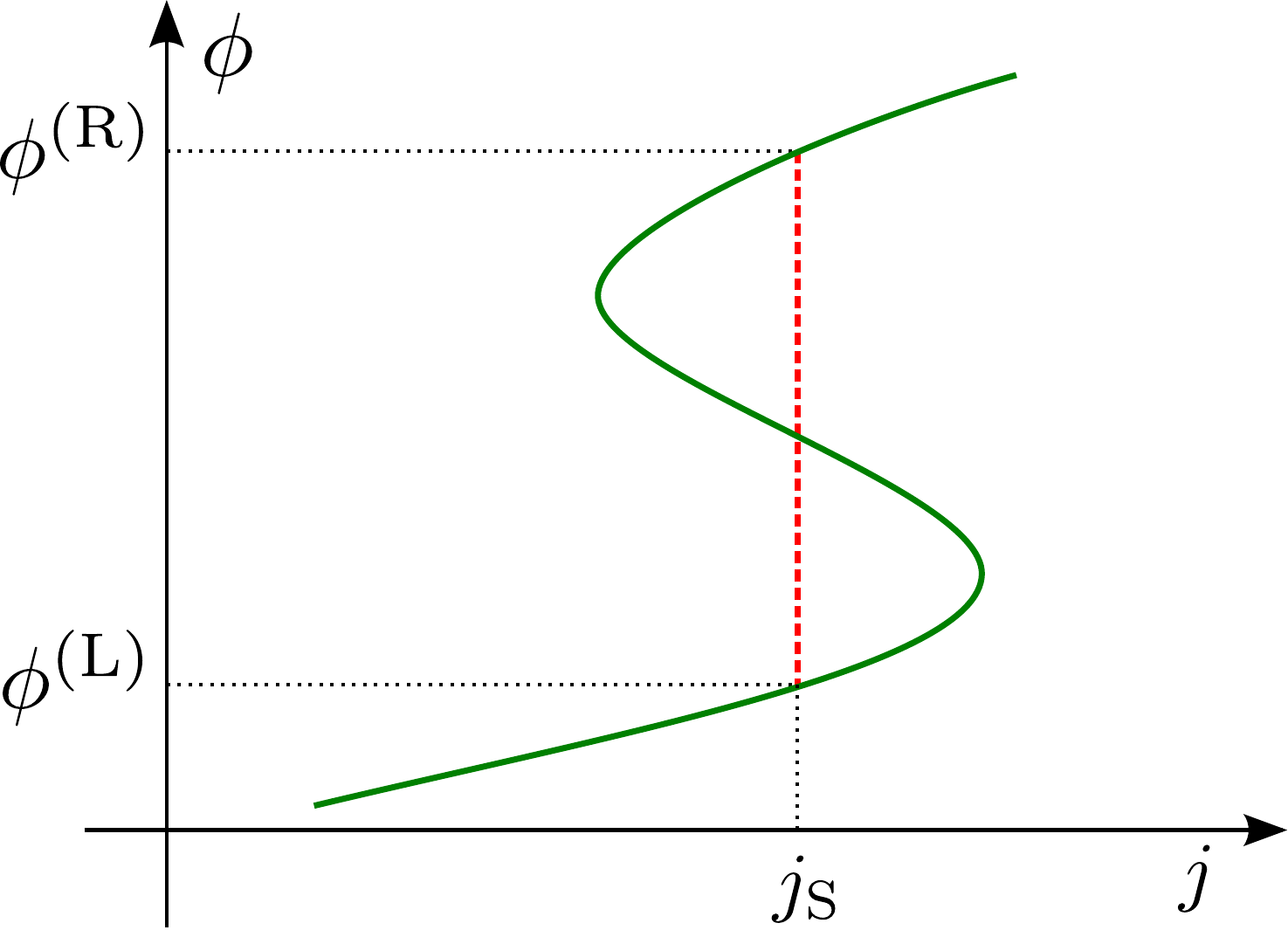}

  {\small (a)}
 \end{minipage}
 \hskip10pt
 \begin{minipage}{0.5\hsize}
  \centering
  \includegraphics[scale=0.4]{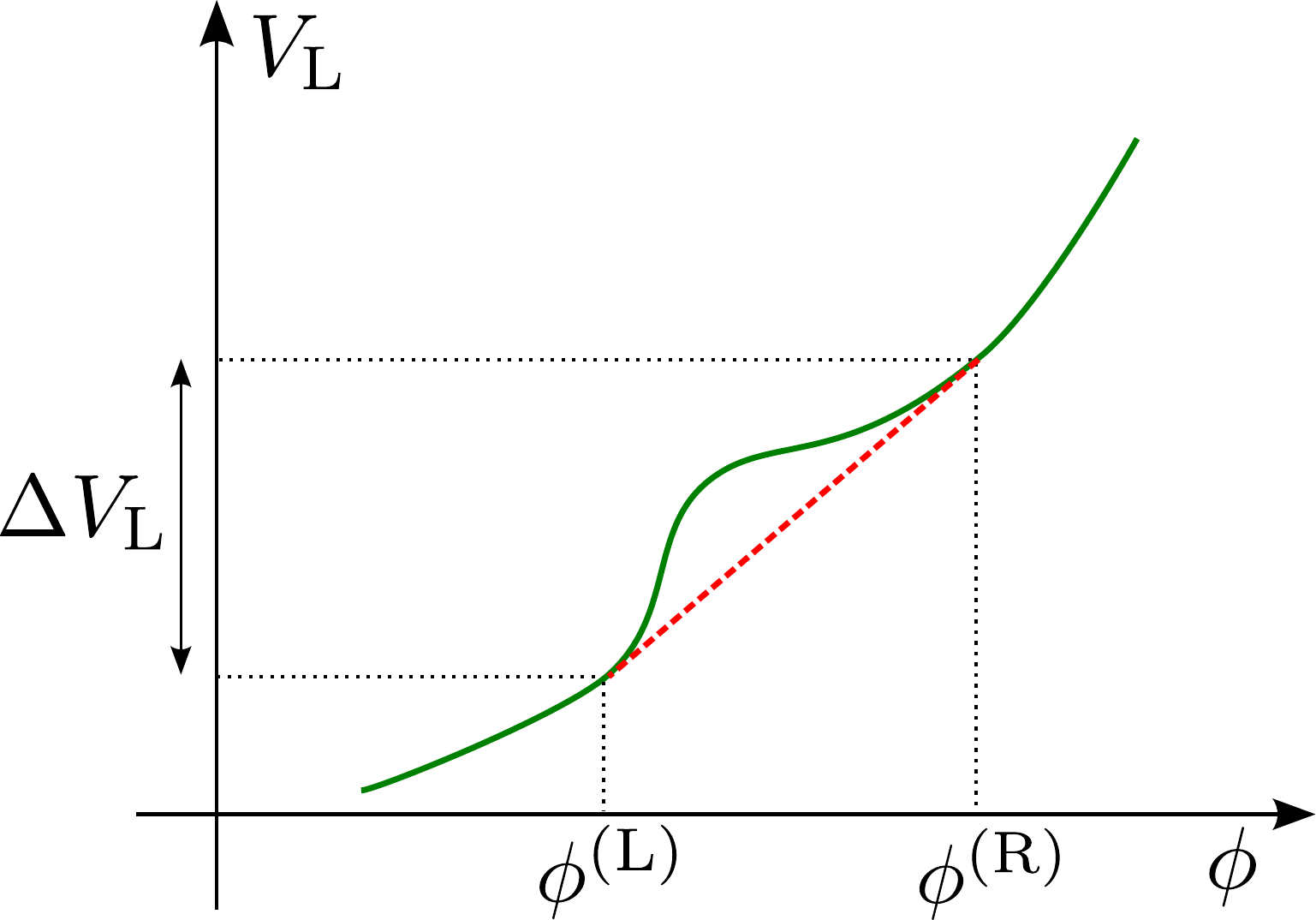} 

  {\small (b)}
 \end{minipage}
\caption{Weak solution of the chiral condensate and the Legendre effective potential $V_L(\phi)$.}
\label{fig:WeakVL}
\end{figure}

As seen previously, because of the translational invariance of the system with respect to $x$, 
the $j$-dependence of $V$ is determined by applying Eq.\,(\ref{eq:V_shift}),
\begin{align}
V( x;t;j)=V( x+j/G_0;t;j=0)-\frac{m_0 j}{G_0}-\frac{j^2}{2G_0}, 
\label{eq:translation}
\end{align}
Thus the free energy and the chiral condensate are calculated 
through the quantities with $j=0$ as follows,
\begin{align}
W(j;t) &= V( x=0;t;j)= V(j/G_0; t;j=0)-\frac{m_0 j}{G_0}-\frac{j^2}{2G_0}, \\
\phi(j;t)& =\frac{1}{G_0} \left[ 
       M(j/G_0;t;j=0)-m_0-j 
    \right].\label{eq:phiofj}
\end{align}
Therefore, we conclude that the chiral condensate function $\phi(j;t)$ has the same structure
of multi-valuedness as the mass function, which is shown 
in Fig.\,\ref{fig:WeakVL}(a).

We have obtained the weak solution of mass function $M(x;t)$, which determines the
weak solution of the chiral condensate $\phi(j;t)$.
The weak solution of the mass function is easily obtained by the equal area rule.
This rule does work as well for the weak solution of the chiral condensate, since the 
difference between functions of $M$ and $G_0 \phi$ 
are well-defined single valued function, $m_0 + j$, 
and it does not contribute to the total area.
We denote the discontinuity of $\phi$ at the singularity by $\phi^{\rm (L)}$ and 
$\phi^{\rm (R)}$, which are values on each side of the singularity, respectively.

The behavior of the Legendre effective potential in the corresponding multi-valued sector is drawn
in  Fig.\,\ref{fig:WeakVL}(b).
Note that between $\phi^{\rm (L)}$ and $\phi^{\rm (R)}$, the function $j(\phi)$
is not monotonic 
and therefore the Legendre effective potential in that region breaks the convexity
(monotonicity of the derivative).
We evaluate the jump of the Legendre effective potential at the singularity ($j=j_{\rm s}$),
\begin{align}
\Delta V_{\rm L} &= V_{\rm L}(\phi^{\rm (L)}) -  V_{\rm L}(\phi^{\rm (R)})
= j _{\rm s}(\phi^{\rm (L)}-\phi^{\rm (R)} ) -(W^{\rm (L)} - W^{\rm (R)}) \notag\\
&= j_{\rm s} (\phi^{\rm (L)}-\phi^{\rm (R)} ) -(V^{\rm (L)} - V^{\rm (R)})
=  j _{\rm s}(\phi^{\rm (L)}-\phi^{\rm (R)} ) ,
\label{eq:RH_VL}
\end{align} 
where we have used 
the continuity of the fermion potential $V$ proved in Eq.\,(\ref{eq:RHV}) which 
assures the continuity of the free energy $W$ at the singularity.

Taking account of the fact that $j$ is equal to the derivative of the Legendre effective
potential at both of $\phi^{\rm (L)}$ and $\phi^{\rm (R)}$, the condition in Eq.\,(\ref{eq:RH_VL})
means that there is a common tangent line between these two points.
Therefore the weak solution of the Legendre effective potential is the envelope 
of it. In other words the weak solution condition automatically convexifies the 
Legendre effective potential.
Any patch working of function $\phi(j)$ in Fig.\,\ref{fig:WeakVL}(a) 
does assure the monotonic change of j as a function of $\phi$. However it is 
not enough to have the convexified effective potential. 
If we do a wrong patch work (wrong placing of $j_{\rm s}$) violating the RH condition, 
then the fermion potential $V$ has a discontinuity, and accordingly
the Legendre effective potential $V_{\rm L}$ is not even uniquely determined
as a function of $\phi$.

In this paper we do not demonstrate to calculate these potentials for 
the zero density NJL model, instead we make plots for the finite density model
in the next section. By examining the plots there, 
readers may easily understand what happens for the 
second order phase transition case of the zero density NJL, since 
it is simpler and more straightforward. 

The automatic convexification proves the greatest feature of the weak solution, 
although it is not so obvious in case of the 
second order phase transition, where the convexified effective potential just shows up as that with 
a flat bottom connecting the symmetry breaking pair of vacua.
In case of the first order phase transition, however, this is crucial, since the convexified 
effective potential correctly selects the globally lowest free energy vacuum even when 
there are meta-stable vacua around.
We will demonstrate it in the next section.

\section{First order phase transition at finite chemical potential}
\begin{figure}[h]
\centering
\includegraphics[width=0.5\hsize]{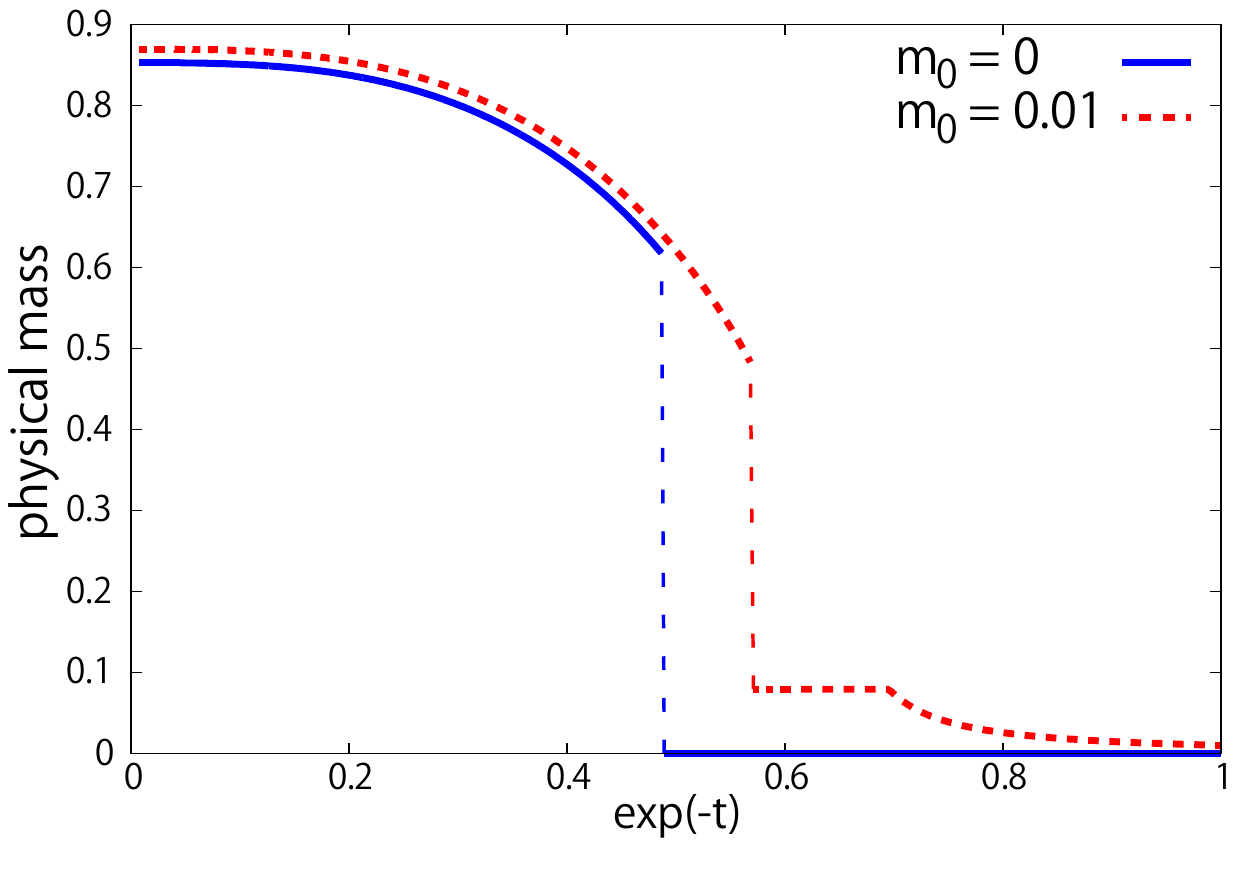}
\caption{RG evolutions of the physical mass at finite density.
}
\label{fig:1st_physmss}
\end{figure}

In this section we investigate the first order phase transition observed 
at finite chemical potential ($\mu\neq 0$).
The first order phase transition is highly non-trivial compared to the second order 
phase transition because the RG evolution of the physical mass has a finite jump 
even with the bare mass $m_0\neq 0$ (as shown in Fig.\,\ref{fig:1st_physmss}).
Therefore the bare mass does not help at all.
Moreover the effective potential has the multi-local minima and most important is whether
the NPRG solution can pick up the physically correct, globally lowest free energy vacuum or not.   
We show that the weak solution at finite chemical potential is uniquely constructed, 
and the effective potential obtained is automatically ``convexified",
which assures the solution is the physically correct lowest free energy vacuum.

We consider the finite density NJL model.
The chemical potential $\mu$ is introduced by adding the term $\mu\bar\psi\gamma_0\psi$ 
to the bare Lagrangian\,(\ref{eq:NJL}).
To make the NPRG equation best appropriate for the future finite temperature calculation,
we use the spacial three dimensional momentum cutoff: $\sum_{i=1}^3 p_i^2\leq\Lambda^2$ to define the 
renormalization scale.


Calculation of $\beta$-function for the Wilsonian effective potential proceeds as follows.
The $\beta$-function is given by the shell mode path integration, 
\begin{equation}
{\partial_t} V(x;t)
=\frac{1}{2 d t} \int_{-\infty }^{\infty } d {p_{\scalebox{0.6}{0}}}  \int\limits_{\Lambda-\Lambda d t \leq |{\bm p}| < \Lambda} \frac{d^3 p}{(2\pi )^4} ~{\rm tr} \log O \ ,
\end{equation}
where the inverse propagator matrix $O$ is given by
\begin{equation}
  O 
  = \left(
    \begin{array}{cc}
      V_{{\psi}^{\rm T} {\psi}} & -i \Slash{p}^{\rm \scalebox{0.6}{T}}-\mu \gamma^{0}+V_{{\psi}^{\rm T} {\bar{\psi}^{\rm T}}} \\[2pt]
     -i \Slash{p} -\mu \gamma^{0}+V_{\bar{\psi} \psi} & V_{\bar{\psi} {\bar{\psi}^{\rm T}}}
    \end{array}
  \right)\ ,\ \ 
V_{{\bar{\psi}^{\rm T}} {\psi}} =  \frac{\delta^2 V}{\delta {\bar{\psi}^{\rm T}} \delta {{\psi}}},
\ etc.
\end{equation}
We omit the large-$N$ non-leading terms (diagonal elements in $O$) and we get
\begin{align}
{\partial_t} V(x;t)
& = \frac{\Lambda^3}{2 \pi^3}  \int_{-\infty }^{\infty } d        
     {p_{\scalebox{0.6}{0}}}~
     \log  \bigl\{ { (p_{\scalebox{0.6}{0}}}-i\mu)^2
        +\Lambda^2+({\partial_x} V)^2   \bigr\} \\
 &= \frac{\Lambda^3}{\pi^2} 
    \left[
      \sqrt{\Lambda^2+M^2}+\left(|\mu|-\sqrt{\Lambda^2+M^2}\right)\cdot
      \Theta\left(|\mu|-\sqrt{\Lambda^2+M^2}\right)
      +C
    \right],
    \label{eq:FatMuNeq0}
\end{align}
where $\Theta(x)$ is the Heaviside step function and $C$ is a divergent constant 
independent of $\mu$, $\Lambda$ and $M$, and thus we ignore it here.
The RGE for the mass function is obtained as 
\begin{equation}
\begin{split}
{\partial_t} M(x;t) &= \frac{\Lambda^3}{2 \pi^3}  
\int_{-\infty }^{\infty } d {p_{\scalebox{0.6}{0}}} 
~ \frac{2 M \partial_x M}{ {p_{\scalebox{0.6}{0}}}^2-2 i \mu {p_{\scalebox{0.6}{0}}} -\mu^2 + \Lambda^2 + M^2   } 
=(\partial_x M)  \frac{\Lambda^3 M \Theta (\sqrt{\Lambda^2+M^2}-\mu)}{ \pi^2 \sqrt{\Lambda^2+M^2} } .
\end{split}
\end{equation}
The translational invariance with respect to $x$ still holds.
Note that due to the change of the cutoff scheme 
the critical coupling constant at vanishing chemical potential changes,
and we now have $G_{\rm c}=2\pi^2/\Lambda_0^2$.

The characteristic curve is given by the following equation of motion,
\begin{align}
\frac{d\bar x (t)}{dt}= 
-\frac{\Lambda^3 M}{\pi^2\sqrt{\Lambda^2+M^2} } 
\ \Theta\left(\sqrt{\Lambda^2+M^2}-\mu\right).
\end{align}
The {\sl particle} motion stops suddenly at some renormalization scale due to the  $\Theta(x)$ 
factor and the smaller the {\sl momentum} $M$, the earlier the {\sl particle} stops. 

\begin{figure}[h]
\centering
\begin{minipage}{0.45\hsize}
\centering
\includegraphics[width=\hsize]{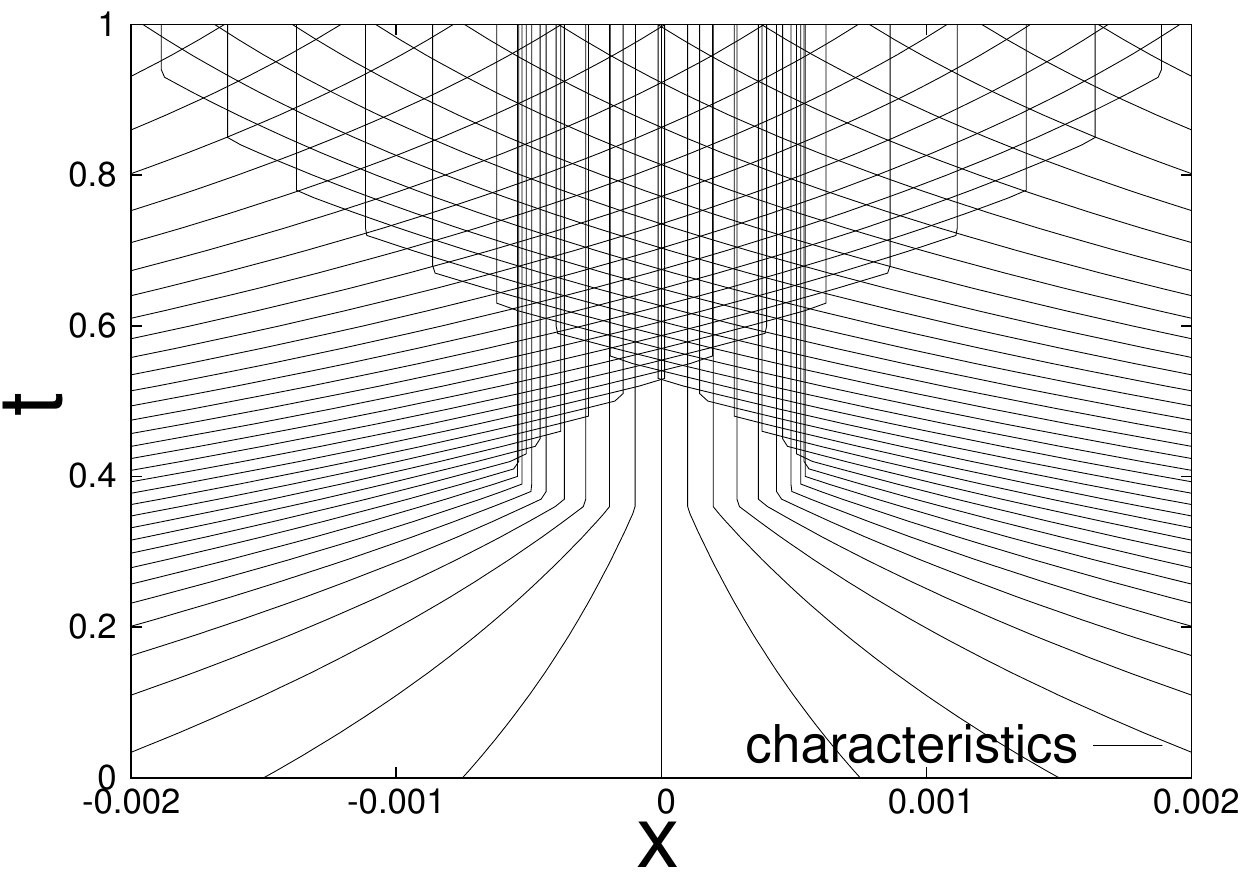}
\\ \hspace{10pt} {\small (a)}
\end{minipage}
\hspace{3pt}
\begin{minipage}{0.45\hsize}
\centering
\includegraphics[width=\hsize]{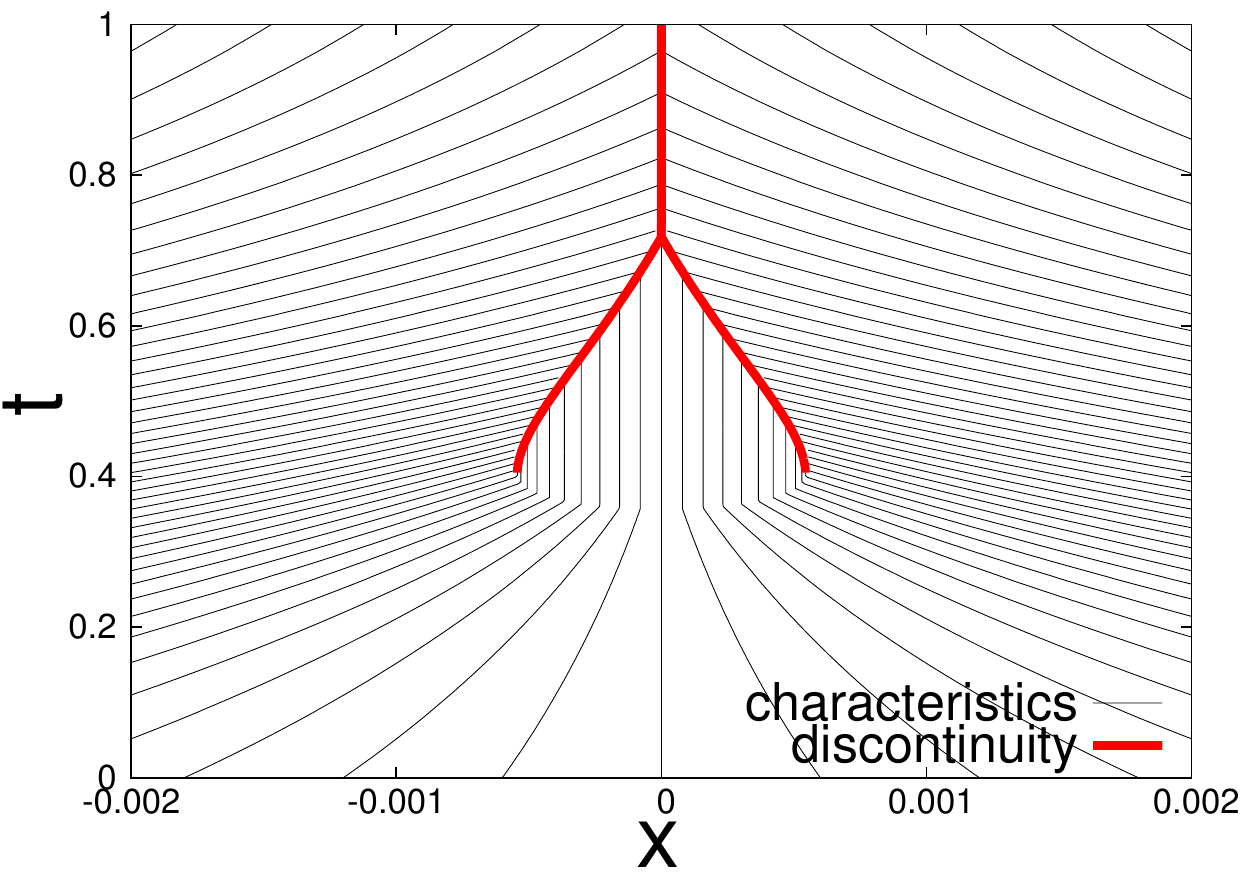}
\\ \hspace{10pt} {\small (b)}
\end{minipage}
\caption{(a) Characteristics. (b) Position of discontinuity determined by the RH condition.}
\label{fig:CC}
\end{figure}

Now we solve the system using the characteristics and construct the 
weak solution in order.
In Fig.\,\ref{fig:CC}, we show the characteristics  for the finite density 
NJL model with $m_0=0$, $\mu=0.7.$ and $G_0=1.7G_{\rm c}$.
The characteristics start folding at $t_{\rm c}=0.41$ at some finite $x$ place, and 
it occurs at two places simultaneously.

This is well seen in Fig.\,\ref{fig:WkatMuNeq0}, where the bare mass
is included to be $m_0=0.01\Lambda$ and we plot the mass function  $M$
in the left most column, the fermion potential $V$ in the mid column and the Legendre
effective potential $V_{\rm L}$ in the right most column respectively.
Note that we set a finite bare mass, and accordingly 
the mass function and the fermion potential
are translated to the negative $x$ side as derived in Eq.(\ref{eq:translation}).
If there is no bare mass at all, the chiral symmetry assures these two singularity
starts symmetrically keeping the odd function property of the mass function.

This situation is quite different from the second order phase transition 
at vanishing density. There, the appearance of the singularity means 
the D$\chi$SB and it occurs at the origin.
Now this first emergence of the folding does not mean the chiral symmetry
breaking, and singularities are pairwisely generated.
It just means the convexity of the Legendre effective potential is broken
at some finite $x$ positions. It should be noted that this is not the emergence of 
the meta stable state yet.

The pair of singularities grows up and we define the weak solution of the mass
function $M$ by using the equal area rule just as before which is depicted by 
the dashed line in the mass function.
As for the fermion potential $V$, the weak solution, which must be continuous, 
just takes the upper envelope. It is proved easily that there is no other way of 
making the continuous function. This selection 
is equivalent to the maximum potential principle, 
and it is related to the {\sl maximal}
action principle in the viewpoint of the kinematical system.
This variational property of solution can be directly related to 
another type of weak solution, the viscosity solution \cite{Crandall:1981,Crandall:1983}, 
which will be treated elsewhere.

As time goes by, the singularity points move towards the symmetry center
at $x_{\rm c}=-m_0/G_0$, which is the origin in case $m_0=0$.
In the mid of this movement, between $t=0.4$ and $0.5$, the three folded
branches at the right-hand side crosses the origin $x=0$. 
This happens exactly when there appears
a meta stable state, a local minimum, if we faithfully adopt all the
multi-valued branches.
This is seen in Eq.(\ref{eq:phiofj}), where in the above situation, 
the source $j$ can reach zero on a local (unused) strong solution $M$, and it means 
the derivative of the Legendre effective potential calculated from the 
multi-valued $M$ vanishes there.  

The pair of singularities reaches the symmetry center $x_{\rm c}$ at $t=0.72$. 
Then the singularities are merged into one singularity on the symmetry center, and 
it does stop and will not move forever.
This is the only way of satisfying the RH condition, which is understandable by the
equal area rule. Here note that there are four regions contributing the area. 
After this merge of pair of singularities, all the functions $M, V, V_{\rm L}$
given by the weak solution are very much like those of the
second order phase transition after the breakdown.
There are no effects of meta stable state near the origin.

In Fig.\,\ref{fig:WkatMuNeq0}, the special time $t=0.5615$ is picked up. This is the 
time of the first order phase transition for these specific parameters of the model.
It is characterized by the time that the right singularity crosses the origin. 
The vanishing derivative point of the Legendre
effective potential jumps from the central well to the right-hand side well
as seen in the corresponding figure (4c).
At this phase transition point the chiral condensate does jump although
the bare mass is included to explicitly break the chiral symmetry, which 
is drawn in Fig.\ref{fig:1st_physmss}.

Thus the bare mass does not help us to avoid the jump of the physical
quantities in $t$-development in the first order phase transition case, 
and even in that case the weak solution successfully describes this jump behavior.
We may in principle regularize this jump by limiting the total degrees of
freedom to be finite. Then there cannot be any singularity in all functions and
the jump becomes just a quick change of the physical value.
In this case not only the second order phase transition but also the first
order phase transition are smoothed out to be the crossover. 

Note that most essential point of these procedures is that the phase transition
is correctly described so that the lowest free energy vacuum is automatically 
selected. This selection is done by the weak solution, not by hand.
The weak solution is unique,  that is, starting with the regular and smooth function 
as the initial condition, developing by time, encountering singularities, 
then it automatically treat the multi-folding structure to pick up a single valued function,
and the resultant solution is physically correct.  

The move of singularities are drawn in Fig.{\ref{fig:CC}, where note that
the symmetry center is the origin ($m_0=0)$. Pairwisely generated singularities move towards 
the origin, merged into one and it stays at the origin.
The characteristics are all moving in to the discontinuity, thus it satisfies 
the so-called entropy condition.

\begin{figure}[htbp]
  \centering
    \begin{tabular}{c}
   \hspace{-12pt}
      \begin{minipage}{0.32\hsize}
        \centering
          \includegraphics[scale=0.8]{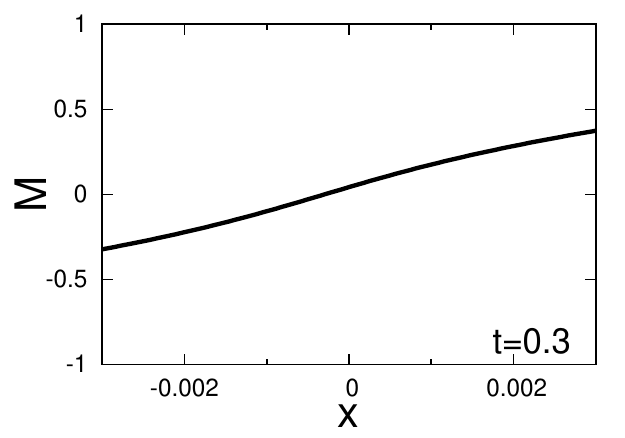}
          
          \vspace{-5pt}
          {\small \hspace{18pt} (1a)} 
      \end{minipage}
      \begin{minipage}{0.32\hsize}
        \centering
          \includegraphics[scale=0.8]{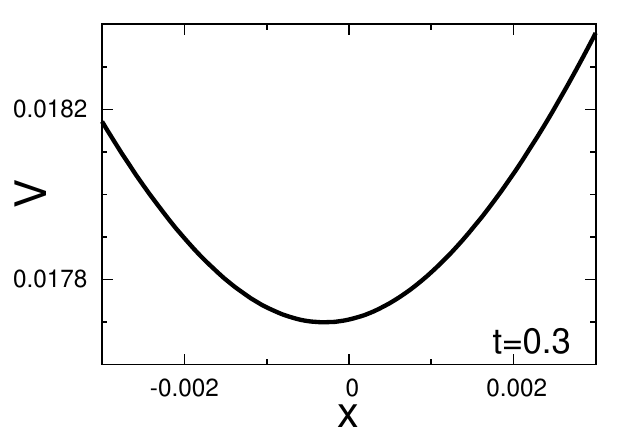}
          
          \vspace{-5pt}
         {\small \hspace{18pt} (1b)} 
      \end{minipage}
      \begin{minipage}{0.32\hsize}
        \centering
          \includegraphics[scale=0.8]{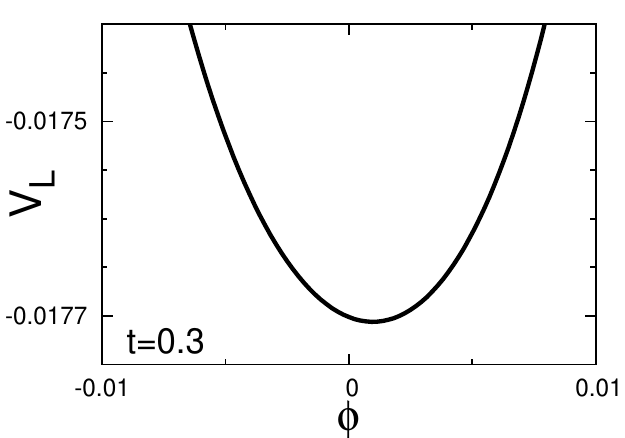}
          
          \vspace{-5pt}
          {\small \hspace{18pt} (1c)} 
      \end{minipage}%
    \end{tabular}      
\vskip 0.15cm
    \begin{tabular}{c}
    \hspace{-12pt}
      \begin{minipage}{0.32\hsize}
        \centering
          \includegraphics[scale=0.8]{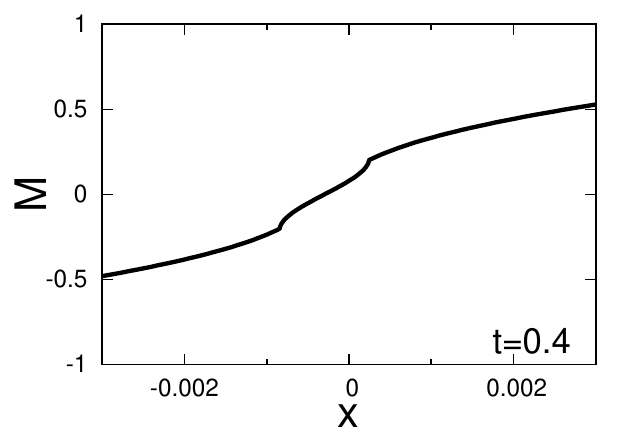}
          
          \vspace{-5pt}
          {\small \hspace{18pt} (2a)} 
      \end{minipage}%
      \hspace{0pt}
      \begin{minipage}{0.32\hsize}
        \centering
          \includegraphics[scale=0.8]{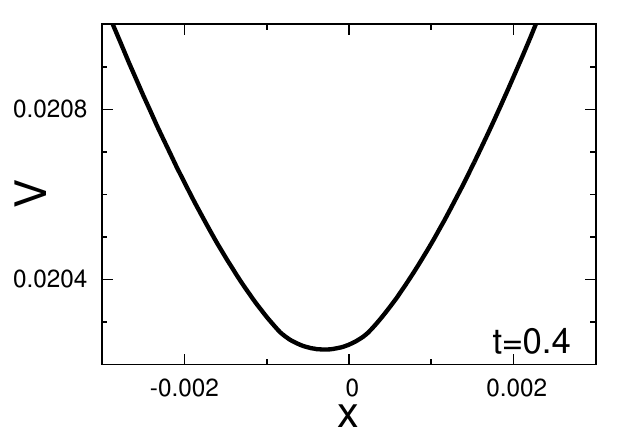}
          
          \vspace{-5pt}
          {\small \hspace{18pt} (2b)} 
      \end{minipage}%
      \hspace{0pt}
      \begin{minipage}{0.32\hsize}
        \centering
          \includegraphics[scale=0.8]{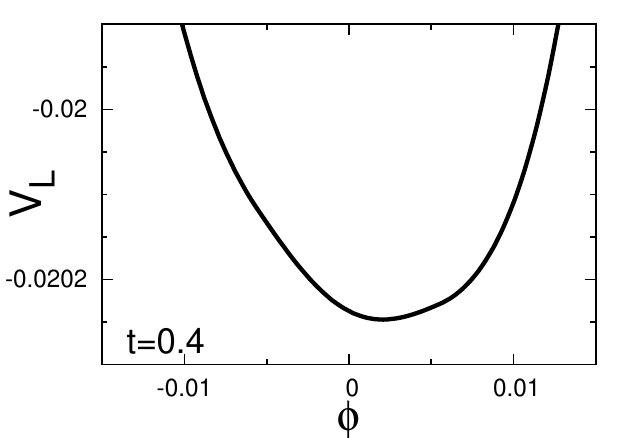}
          
          \vspace{-5pt}
          {\small \hspace{18pt} (2c)} 
      \end{minipage}%
    \end{tabular}
\vskip 0.15cm
    \begin{tabular}{c}
    \hspace{-12pt}
      \begin{minipage}{0.32\hsize}
        \centering
          \includegraphics[scale=0.8]{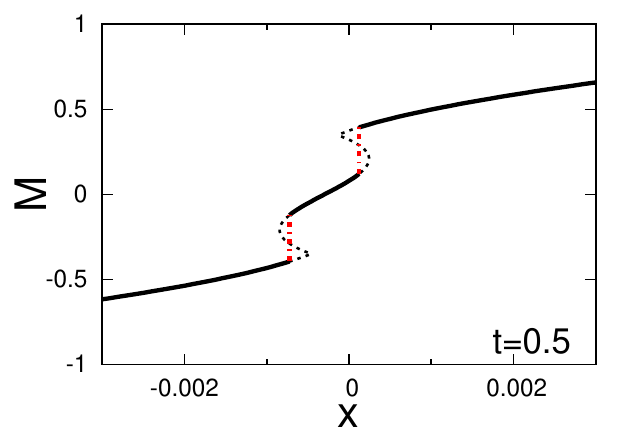}
          
          \vspace{-5pt}
          {\small \hspace{18pt} (3a)} 
      \end{minipage}%
      \hspace{0pt}
      \begin{minipage}{0.32\hsize}
        \centering
          \includegraphics[scale=0.8]{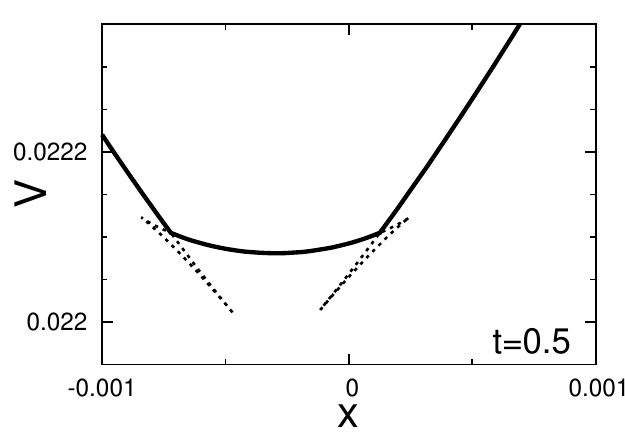}
          
          \vspace{-5pt}          
          {\small \hspace{18pt} (3b)} 
      \end{minipage}%
      \hspace{0pt}
      \begin{minipage}{0.32\hsize}
        \centering
          \includegraphics[scale=0.8]{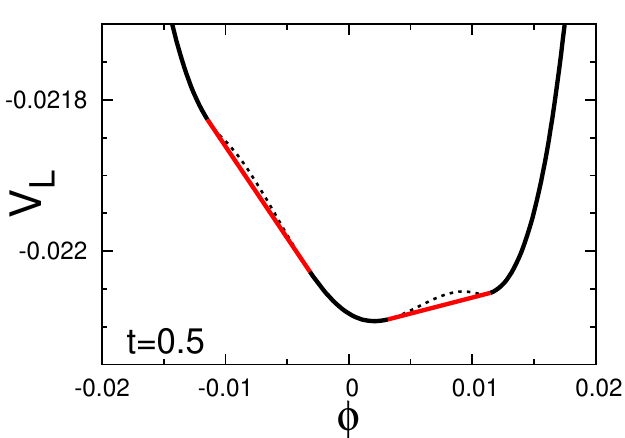}
          
          \vspace{-5pt}          
          {\small \hspace{18pt} (3c)} 
      \end{minipage}%
    \end{tabular}
\vskip 0.15cm
    \begin{tabular}{c}
    \hspace{-12pt}
      \begin{minipage}{0.32\hsize}
       \centering
          \includegraphics[scale=0.8]{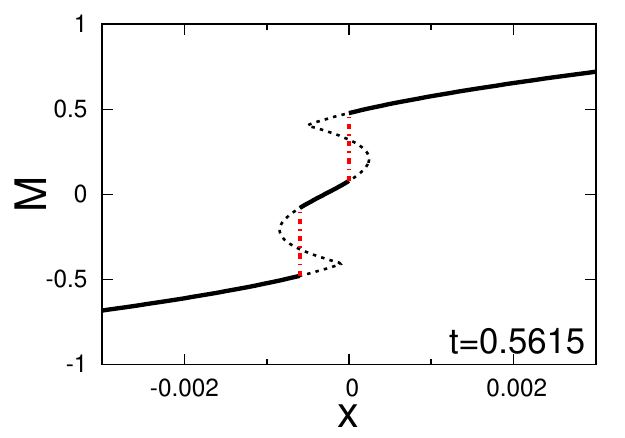}
          
          \vspace{-5pt}          
          {\small \hspace{18pt} (4a) }
      \end{minipage}%
      \hspace{0pt}
      \begin{minipage}{0.32\hsize}
       \centering
          \includegraphics[scale=0.8]{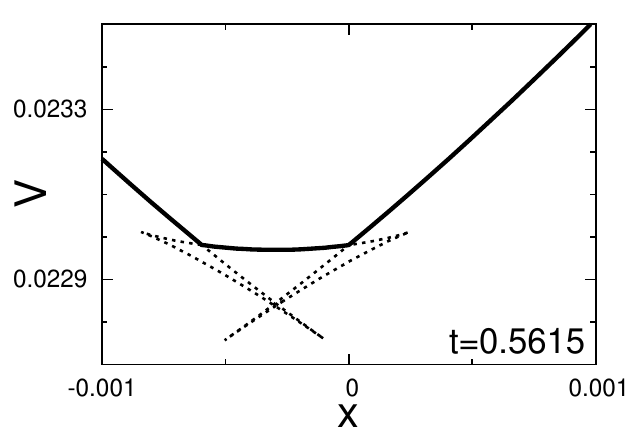}
          
          \vspace{-5pt}          
          {\small \hspace{18pt} (4b)} 
      \end{minipage}%
      \hspace{0pt}
      \begin{minipage}{0.32\hsize}
        \centering
          \includegraphics[scale=0.8]{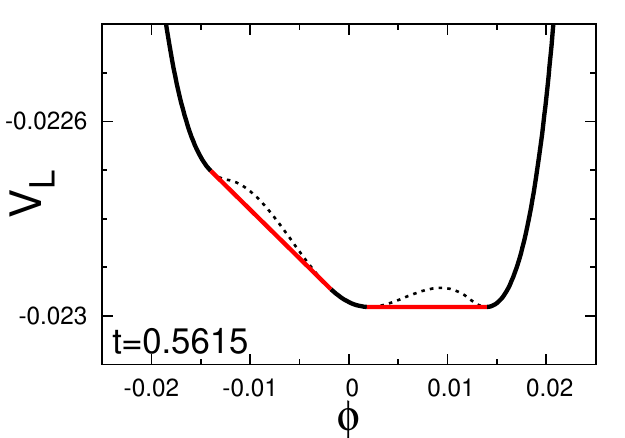}
          
          \vspace{-5pt}          
          {\small \hspace{18pt} (4c)} 
      \end{minipage}%
    \end{tabular}
\vskip 0.15cm
    \begin{tabular}{c}
    \hspace{-12pt}
      \begin{minipage}{0.32\hsize}
        \centering
          \includegraphics[scale=0.8]{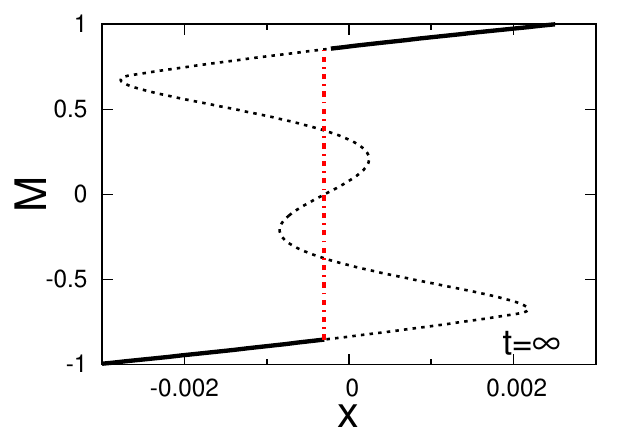}
          
          \vspace{-5pt}          
          {\small \hspace{18pt} (5a)} 
      \end{minipage}%
      \hspace{0pt}
      \begin{minipage}{0.32\hsize}
        \centering
          \includegraphics[scale=0.8]{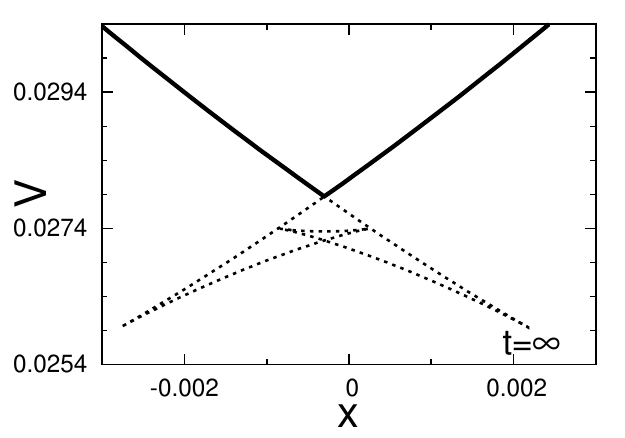}
          
          \vspace{-5pt}          
          {\small \hspace{18pt} (5b)} 
      \end{minipage}%
      \hspace{0pt}
      \begin{minipage}{0.32\hsize}
        \centering
          \includegraphics[scale=0.8]{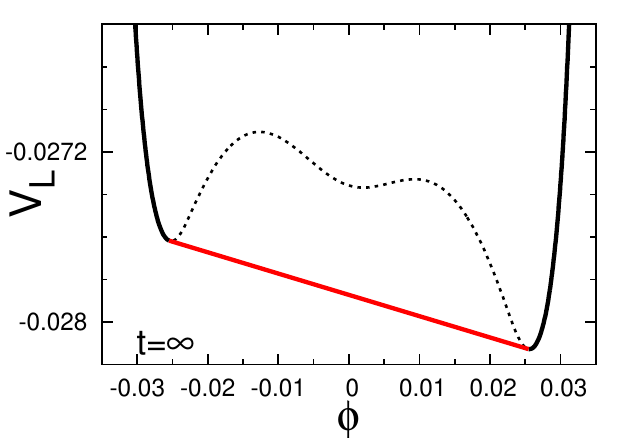}
          
          \vspace{-5pt}          
          {\small \hspace{18pt} (5c)} 
      \end{minipage}%
    \end{tabular}
\caption{RG evolution of physical quantities and the weak solution for the finite density NJL 
with non-zero bare mass 
($G_0=1.7 G_{\rm c}$, $m_0=0.01\Lambda_0$, $\mu=0.7$, 
$~t=0.3$, $0.4$, $0.5$, $0.5615$, $\infty$). 
 (a) The mass function $M(x)$. (b) The fermion potential $V(x)$. 
(c) The Legendre effective potential $V_{\rm L}(\phi)$.
The solid lines denote the weak solution, and the dashed lines denote the local strong solutions 
discarded. 
The straight solid line in (c) corresponds to the jump in $M$ and is actually the envelope.
} 
\label{fig:WkatMuNeq0}
\end{figure}

 \section{Summary}

In this paper, we have introduced the weak solution to define the singular 
D$\chi$SB solution of NPRG equation that can predict physical quantities 
such as the physical quark mass and the chiral condensate.
The weak solution satisfies the integral form of the partial differential equation.
Specifically we have evaluated the weak solution of the large-$N$ NPRG equation 
of the NJL model 
for the mass function which is the first derivative of the fermion potential 
with respect to the scalar bilinear-fermion field $\bar\psi\psi$.
Also we applied our method to the finite density NJL model where the first
order phase transition appears to give highly non-trivial situation compared 
to the second order phase transition case of vanishing density.

There is no global solution for our NPRG equation unless we do not consider the
weak solution, as far as it describes the D$\chi$SB due to the quantum
corrections, that is, the mass function must encounter singularities at a finite
scale.
The weak solution helped us perfectly to define the global solution up to the
infrared limit of the infinite time where we can evaluate physical quantities.
The weak solution is uniquely obtained, and there is no ambiguity or 
approximation to get it. 

The Rankine-Hugoniot condition assuring the weak solution is modified 
into the equal area rule, by which we can easily obtain the weak solution
from the multi-folded local strong solutions.
However here must be some questions.
Since our partial differential equation is the renormalization group equation
and must be solved with the initial condition only, that is, the effective 
action at a time. After it has the discontinuity, it does not know the discarded
strong solutions behind. The effective interactions must determine 
its development with the information given by the weak solution only.
Therefore there can be no notion of ``area'' referring to all multi-valued leaves.

This is true. We should note again that the equal area rule is a result given 
by the weak solution, not an assumption to determine the weak solution.
We just use it in order to have the weak solution easily.
In fact, the Rankine-Hugoniot condition is written down as continuity of $V$ or
conservation law for $M$, and it is the local condition without referring to the
strong solutions behind and it can determine the solution without
recourse to the discarded strong solutions.
Thus the weak solution with singularity can develop by itself
without any additional information.

The weak solution defined in this paper has a  perfect feature to describe
the physically correct vacuum even when there are multi meta stable
local minima given by all the local strong solutions 
in the Legendre effective potential. The basic logic to 
achieve this is that the weak solution successfully convexifies the Legendre
effective potential, and it assures that the lowest global minimum
determines the vacuum.

In this paper we have worked our only the NJL type models with large-$N$
leading approximation. We can go ahead to define the weak solution of 
the NPRG equation for gauge theories even with non-ladder 
(large-$N$ non-leading) effects. For example, the large-$N$
leading NPRG partial differential equation is defined by,
\begin{align}
 \partial_t V(x;t) &=
 \frac{\Lambda^4}{4\pi^2}\ln\left[1+
   \Lambda^{-2}\left(
      \partial_x V+(3+\xi)\frac{ C_2\pi \alpha_{\rm s}}{\Lambda^2} \, x
               \right)^2
   \right]\ ,
\end{align}
which has the same type of PDE analyzed in this paper with 
translationally non-invariant {\sl Hamiltonian} $F$.
Also the extension including large-$N$ non-leading diagrams has been done
to give
\begin{align}
\begin{split}
 \partial_t V(x;t)&= 
 \frac{\Lambda^4}{4\pi^2}\log\left[1+\frac{B^2}{\Lambda^2}\right]
   +\frac{\Lambda^4}{8\pi^2}
     \log\left[
      \frac{\Lambda^2+B^2}{\Lambda^2+{\color{black} (\partial_x V)^2}}
      +\frac{3\Lambda^2G^2}{(\Lambda^2+{\color{black} (\partial_x V)^2})^2}
     \right] \\
   &\quad +\frac{\Lambda^4}{4\pi^2}
     \log\left[
      1+\xi\frac{\partial_x V\, G}{\Lambda^2+{\color{black} (\partial_x V)^2}}
     \right],
\end{split}
\end{align}
where $B=\partial_x V+2C_2\pi\alpha_{\rm s}x/\Lambda^2$ 
and $G=2 C_2 \pi\alpha_{\rm s}x/\Lambda^2$\cite{Aoki:2012mj,Aoki:2000dh}. 
This is also manageable by our weak solution method.
Application to gauge theories are reported in the forthcoming paper.

\begin{figure}[h]
\centering
\includegraphics[scale=0.4]{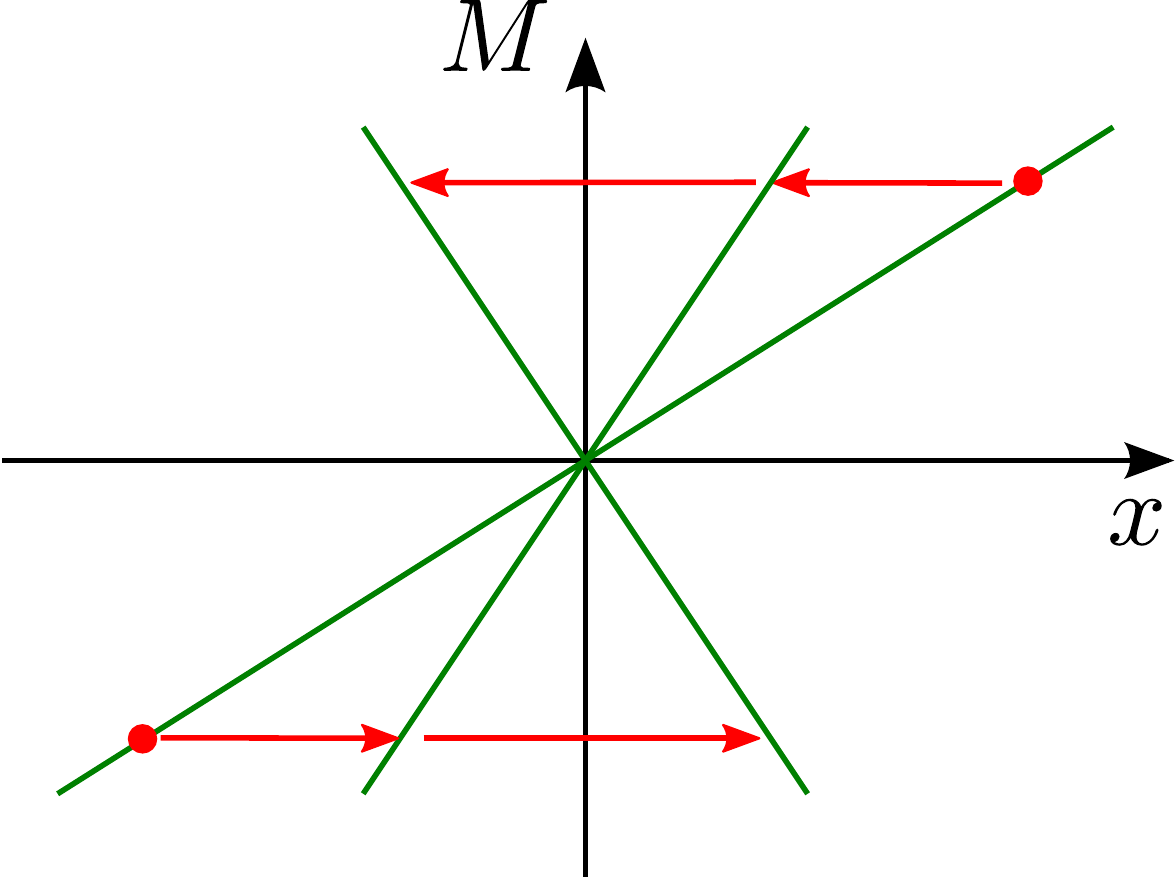}
\caption{Central engine of D$\chi$SB.}
\label{fig:FreeStringMotion}
\end{figure}

Finally we mention about the driving engine of the D$\chi$SB 
taking the NJL model as an example.
Let us see the neighborhood of the origin, $M \simeq 0$, in Eq.\,(\ref{eq:NJL_Fterm}) 
and we have the approximated
{\sl Hamiltonian} $F$ as follows,
\begin{align}
F(M,x;t) = - \frac{\Lambda^4}{4\pi^2}\log\left(1+\frac{M^2}{\Lambda^2}\right)
\simeq - \frac{\Lambda^2}{4\pi^2} M^2 = - \frac{\Lambda_0^2 e^{-2t}}{4\pi^2} M^2\ .
\end{align}
Therefore this {\sl particle} is a usual classical particle but with negative and
time varying {\sl mass} of ${2\pi^2 e^{2t}}/{\Lambda_0^2}$ \cite{Aoki:2012mj} .
The {\sl particle} motion becomes gradually slow due to the exponential 
increase of mass. 
The string near the origin is approximated by a straight line.
The velocity is proportional to the momentum, with negative sign though, and
the string near the origin just turns counter-clockwisely 
until it stops (Fig.\,\ref{fig:FreeStringMotion}).
This is the central engine of generating the singularity 
and of bringing the D$\chi$SB.

Note that the D$\chi$SB occurs exactly when the string at the origin becomes vertical.
Then the derivative of the mass function $M(x;t_{\rm c})$ with respect to $x$
becomes divergent at the origin. 
We denote this slope by $\tilde{G}(t)$ and it is evaluated as
\begin{align}
\tilde{G}(t) = \frac{M(\bar{x}(t);t)}{\bar{x}(t)} e^{-2t}\Lambda_0^2\ .
\end{align}
The RG equation for $\tilde{G}(t)$ reads
\begin{align}
\frac{d}{dt}\tilde{G(t)}=-\left(2+\frac{1}{\bar{x}}\frac{d\bar{x}}{dt}\right)
\tilde{G}
=-\left(2+\frac{1}{\bar{x}}\frac{dF(M,x;t)}{dM}\right)\tilde{G}
= -2\tilde{G}+ \frac{1}{2\pi^2}\tilde{G}^2.
\end{align}
This is nothing but the RG equation of the four-fermi interactions
obtained in Eq.(\ref{eq:NJLbeta}), whose blow up solution encounters the singularity at 
$t_{\rm c}$.
However if we work with the motion of string, the verticality of string 
does not mean anything singular; it is just vertical. 

Some authors define inverse four-fermi coupling constant 
$g(t)\equiv 1/\tilde{G(t)}$, 
which satisfies the following RG equation\cite{Kodama:1999if},
\begin{align}
\frac{d}{dt}g(t) = + 2g - \frac{1}{2\pi^2}.
\end{align}
This equation has the global solution
\begin{align}
g(t) = \frac{1}{4\pi^2} + \left(g(0) - \frac{1}{4\pi^2}\right)e^{2t}\ .
\end{align}
In the strong phase ($g < 1/4\pi^2$), 
$g(t)$ crosses the origin at $t_{\rm c}$ and goes to the negative
region without any singularity. It has a global solution up to $t=\infty$.
This tricky procedure is not authorized by itself. However, it is perfectly right
if we consider the string motion and just use the inverse slope instead of
the  slope to describe the string near the origin.
Sometimes in case of finite density media, 
the inverse coupling constant $g(t)$ goes to the negative region 
and returns back to the positive region again. Even in such case, it is understood
to have given the right result, that is, the chiral symmetry is not broken 
in the macro effective theory.

In gauge theories the same engine does exist but there is a little difference. 
The gluons are massless and have no intrinsic scale, 
and the engine does not stop at the infrared.
Therefore, the string rotates infinitely around the origin, which corresponds
to the infinitely many unphysical solutions encountered in the SD type analyses.  

This singularity generation mechanism quite resembles to the generation of a
shock and its expansion procedure described by 
the non-linear wave equation of the Burgers' \cite{Burgers:1940}.
Also our NJL weak solution satisfies the so-called entropy condition and 
it is understandable as a physical shock. 
We have done this argument using the NJL ($\mu=0$) model near the origin.
However it can be applied as well to any neighborhood of the singularity
birth place.
Thus we demonstrated that the D$\chi$SB is the spontaneous generation 
and growth of a shock describable by the weak solution of PDE.

The generation and growth of a non-moving shock describes the ``continuous''
behavior in the second order phase transition. 
On the other hand, a moving shock realizes the ``jump'' behavior of the first
order phase transition.

We stress again that this is the first demonstration and successful
application of the weak solution for 
the non-perturbative renormalization group equation.
Why does it work so well? One intuitive plausible argument follows.
The definition of the weak solution is given by some integration equality
without recourse to the singularities of the target function.
Our target functions are effective interactions which are to be path integrated
with operators. Therefore such functions do not have to be regular as they are, 
and it may be enough that they satisfy necessary equations as a form of 
integration convoluted with smooth and bounded test functions.

In the introduction we claimed that encountering the singularity in the mid of
renormalization group solution is 
inevitable in any, approximated or not, model of the D$\chi$SB. 
If the PDE system here is the continuum approximation of the originally 
discrete many body system, then we may add higher derivative terms, for
example, the dissipation term, as improvement of approximation, and
such dissipation may regulate the singularity. This is well known in case of the
Burgers' equation. In our PDE of the non-perturbative renormalization
group equation, though, there may appear higher derivative contribution
in improving the approximation, but
it cannot help at all, because our singularity is intrinsic and
is not a matter of approximation. 

There may be another way out by changing the system drastically. 
If we invent some regularization method to 
limit the number of degrees of freedom $N$ to be finite, 
the spontaneous symmetry breaking cannot occur and all effective 
interactions and thermodynamic functions are singularity free.
Then, after all calculations, we take the infinite $N$ limit to get the
physical results. 
However there has not been known any good method of realizing finite degrees of
freedom regularization in formulating the non-perturbative renormalization 
group equation.
We have found a way of limiting the {\sl depth} of quantum corrections, and 
the D$\chi$SB is regularized. After obtaining the
thermodynamic potentials we take the infinite {\sl depth} limit, and finally
reach the right functions with singularities. We will report it in a separate article. 

\subsection*{Acknowledgements}
The authors greatly appreciate stimulating discussion with Yasuhiro Fujii, 
and also helpful lectures by Prof. Akitaka Matsumura 
who told us how to define and construct the weak solution, 
which initiated our work in this paper.

\bibliographystyle{ptephy}
\bibliography{FRGv0}

\end{document}